\documentclass[twocolumn]{aastex62}

\submitjournal{ApJ}

\shorttitle{Atmospheric Dynamics on Terrestrial Planets}
\shortauthors{I. Guendelman and Y. Kaspi}

\begin{document}

%\title{An Example Article using \aastex v6.2\footnote{Released on January, 8th, 2018}}
\title{Atmospheric Dynamics on Terrestrial Planets: The Seasonal Response to changes in Orbital, Rotational and Radiative Timescales.}

\correspondingauthor{Ilai Guendelman}
\email{ilai.guendelman@weizmann.ac.il}

\author{Ilai Guendelman}
\affiliation{Department of Earth and Planetary Sciences, Weizmann Institute of Science \\
234 Herzl st., 76100 \\
Rehovot, Israel}

\author{Yohai Kaspi}
\affiliation{Department of Earth and Planetary Sciences, Weizmann Institute of Science \\
234 Herzl st., 76100 \\
Rehovot, Israel}
\collaboration{}

\begin{abstract}
Thousands of exoplanets have been detected to date, and with future planned missions this tally will increase. Understanding the climate dependence on the planetary parameters is vital for the study of terrestrial exoplanet habitability. Using an idealized general circulation model with a seasonal cycle, we study the seasonal response of the surface temperature and Hadley circulation to changes in the orbital, rotational and radiative timescales. Analyzing the climate's seasonal response to variations in these timescales, we find a regime transition between planets controlled by the annual mean insolation to planets controlled by the seasonal variability depending on the relation between the length of the orbital period, obliquity and radiative timescale. Consequently, planets with obliquity greater than $54^{\circ}$ and short orbital period will have a minimum surface temperature at the equator. We also show that in specific configurations, mainly high atmospheric mass and short orbital periods, high obliquity planets can still have an equable climate. Based on the model results, we suggest an empirical power law for the ascending and descending branches of the Hadley circulation and its strength. These power laws show that the Hadley circulation becomes wider and stronger by increasing the obliquity and orbital period or by decreasing the atmospheric mass and rotation rate. Consistent with previous studies, we show that the rotation rate plays an essential role in dictating the width of the Hadley circulation.
\end{abstract}

\keywords{atmospheric circulation - terrestrial planets }


\section{Introduction} 
\label{sec:intro}
Thousands of planets outside of the solar system have been detected in the last two decades. These planets vary in their mass, radius, orbital period, eccentricity, solar constant, and more. In this variety of planets, terrestrial planets within their host star habitable zone, are of particular interest due to their potential to harbor life on their surface. Planetary and atmospheric characteristics can change the planet habitability potential \citep[e.g.,][]{spiegel2009habitable,kopparapu2013habitable,kopparapu2017habitable,vladilo2013habitable,armstrong2014effects}, and will have a large impact on the planet's atmospheric dynamics \citep[e.g.,][]{ferreira2014climate, kaspi2015atmospheric, linsenmeier2015climate, chemke2017dynamics, penn2018atmospheric}. Studying the atmospheric dynamics dependencies on planetary parameters, will not only improve the understanding of atmospheric physics but may give rise to possible future observables from which we can learn about the planet and its atmosphere characteristics.  

This study focuses mainly on seasonal variations of climate. We focus on three timescales that define the climate's seasonal variability: the rotational, orbital, and radiative timescales. This is done through a series of simulations varying the rotation rate ($\Omega$) and three other parameters that are closely related to the radiative forcing, the obliquity ($\gamma$), orbital period ($\omega$) and atmospheric surface pressure ($p_s$). We examine the zonal mean climate, more specifically the zonal mean meridional circulation and surface temperature dependence on the parameters mentioned above. The mean meridional circulation is one of the main circulation features of a terrestrial atmospheres \citep{vallis2017atmospheric}, which on Earth, strongly relates to the water cycle and on Titan to the methane cycle \citep[e.g.,][]{mitchell2006dynamics}. On Earth, this meridional circulation in the tropics is dominated by the thermally driven Hadley cell up to the subtropics. Beyond that, the meridional circulation is driven by the turbulence forming the Ferrel cell \citep{vallis2017atmospheric}. The zonal mean meridional circulation, is described using the zonal mean meridional streamfunction, $\psi =2\pi a\int\overline{v}\cos\phi d\sigma$ where $a$ is the planetary radius, $\overline{v}$ is the zonal mean meridional wind, $\phi$ is latitude and $\sigma=p/p_s$ is the vertical pressure coordinate normalized by the surface pressure.  

The atmospheric circulation dependence on the rotation rate is a result of the Coriolis acceleration importance in the momentum balance. Fast rotation rate results in a multicellular structure of the streamfunction together with multiple jets, while as the rotation rate is slowed down, the cell number decreases and the Hadley cell widens and strengthens \citep[e.g.,][]{walker2006eddy, kaspi2015atmospheric, chemke2015latitudinal, chemke2015poleward}. Slowing down the rotation rate results also in a weaker meridional temperature gradient. This is due to the eddy scale dependence on rotation rate, where eddies become smaller as the rotation rate is increased \citep{walker2006eddy, kaspi2015atmospheric}, resulting in a less efficient meridional heat transport that increases the meridional temperature gradient.

\begin{figure}[htb!]
\centerline{\includegraphics[height=4cm]{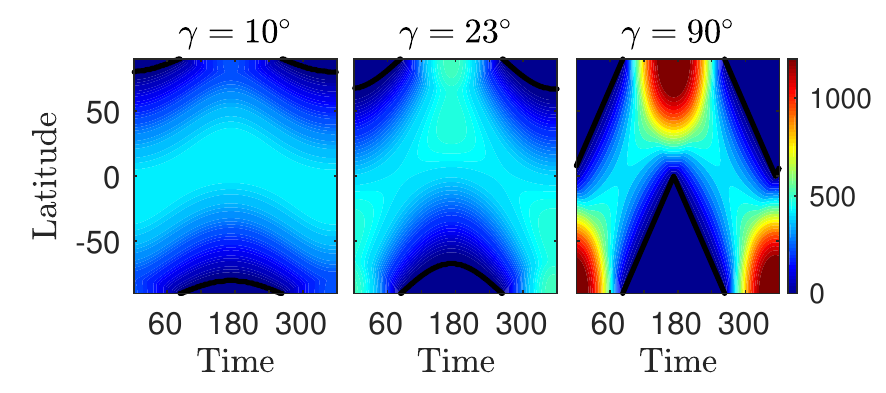}}%{fig_1/pdf_fig/Fig_ob_exam.pdf}}
\caption{The diurnal mean insolation for different values of obliquity, $10^{\circ},23^{\circ}$ and $90^{\circ}$. Black contours represent the beginning of zero insolation zone.}
\label{fig:figtwo}
\end{figure}

The other three parameters that this study focuses on are closely related to the radiative forcing. The obliquity ($\gamma$), determines the seasonality of the insolation. Non-zero obliquity introduces seasonality to the insolation, and increasing the obliquity towards $90^{\circ}$, shifts poleward the latitude of maximum insolation and increases the insolation meridional gradient (Fig. \ref{fig:figtwo}). At obliquity $90^{\circ}$, for every degree the planet moves in its orbit the latitude of zero insolation (black contours in Fig. \ref{fig:figtwo}) also moves a degree, which explains the linearity seen in the zero insolation line in Fig.~\ref{fig:figtwo}c. Several studies used different types of general circulation models (GCMs) to study the circulation and habitability dependence on the obliquity. There is a general agreement between the models that seasonal variability increases by increasing the obliquity. Most of these studies though focus mainly on the effect of obliquity on the planetary habitability \citep{williams1997habitable, spiegel2009habitable,armstrong2014effects, ferreira2014climate, linsenmeier2015climate, wang2016effects, nowajewski2018atmos}. \cite{mitchell2014seas} studied the seasonality effect on climate, using an idealized parameterization for the seasonality and radiative timescale. In contrast, in this study we examine the climate seasonality response to different physical parameters. This study also provides a systematic study of the seasonal cycle climate dynamical response to changes in the obliquity, and the obliquity effect in a larger parameter space. This study can be viewed as an expansion of \cite{kaspi2015atmospheric} perpetual equinox study to include the effects of seasonality. 

\begin{figure}[htb!]
\centerline{\includegraphics[height=4cm]{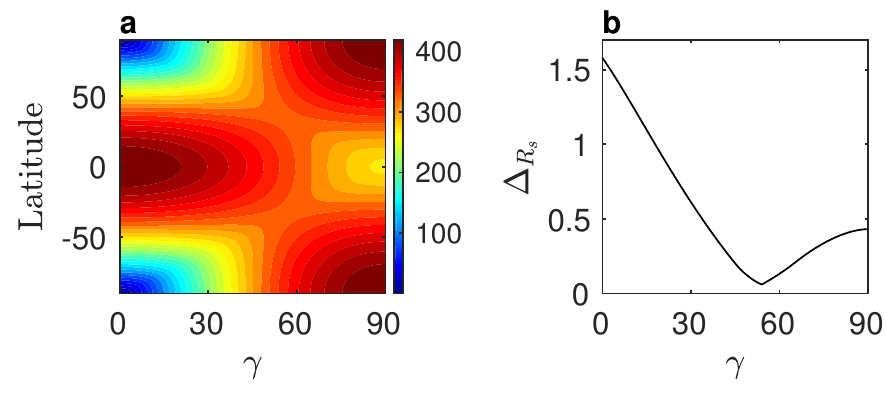}}%{fig_1/pdf_fig/Fig_ob_mean.pdf}}
\caption{The yearly mean insolation and insolation gradient depndence on the obliquity. Panel a shows the yearly mean insolation as a function of latitude and obliquity ($\gamma$) \citep[similar to Figure 2 in][]{linsenmeier2015climate}. Panel b shows the normalized yearly mean insolation gradient, $\Delta_{\mathit{R_s}}=\frac{\rm{max}(\mathit{R_s})-\rm{min}(\mathit{R_s})}{\rm{mean}(\mathit{R_s})}$ as a function of obliquity.}
\label{fig:figthree}
\end{figure}

Introducing seasonality to the climate system requires taking into account different timescales. The orbital period is a natural timescale of the seasonal cycle, dictating the time over which radiative changes take place. The longer the orbital period is, radiative changes occur over more extended periods giving the atmosphere longer time to adjust. As radiative changes take place over shorter periods (short orbital periods), a point where the radiative forcing is effectively the annual mean one (Fig.~\ref{fig:figthree}) is reached. Note, that although the orbital period is coupled to the distance from the host star and thus to the stellar flux reaching to the planet, in this study the stellar flux is kept constant while the orbital period varies. This is analogous to studying similar planets that orbit stars with different mass and thus different luminosity. Due to the simplified radiation scheme in our model, effects related to the spectral distribution of the insolation are not taken into account, although it can have some effect the resulting climate \citep{godlot2015climate, shields2016stars}. 

Another important timescale is the atmospheric radiative timescale, meaning, the time that takes the atmosphere to adjust to radiative changes. This timescale depends on the atmospheric mass per unit area, defined to be $m=p_s/g$ \citep{hartmann2015global}, where $p_s$ is the surface pressure and $g$ is the surface gravity. The radiative timescale is defined as
\begin{equation}
\label{eq:radts}
\tau=\tau_{IR}\frac{c_p p_s}{4g\sigma_b T_e^3},
\end{equation}
where $\tau_{IR}$ is the infrared optical depth of the atmosphere, $c_p$ is the heat capacity of the air, $\sigma_b$ is the Stefan-Boltzmann constant and $T_e$ is the radiative equilibrium temperature. The radiative timescale is linearly proportional to the atmospheric mass (Equation \ref{eq:radts}), meaning that as the atmospheric mass is increased, the atmosphere needs a longer time to adjust to the changes in the radiative forcing. The radiative timescale has a strong dependence on the planetary equilibrium temperature $T_e$, yet as it is not an input parameter of the model and in order to avoid additional complexities as shown in \cite{kaspi2015atmospheric}, we focus on varying the radiative timescale by changing only the atmospheric mass. 

Increasing the atmospheric mass expands the outer edge of the habitable zone due to the increase in surface temperatures with the atmospheric mass \citep[e.g.,][]{wordsworth2010gliese, wordsworth2011gliese, vladilo2013habitable}. Increasing the atmospheric mass also results in flattening of the meridional temperature gradient and lowering of the troposphere height \citep{goldblatt2009nitrogen, kaspi2015atmospheric, chemke2016thermodynamic, chemke2017dynamics}. Several explanations for this increase in the surface temperature have been given; first, increasing the atmospheric mass results in enhancement of the greenhouse effect by pressure broadening of absorption lines overcoming the enhancement of Rayleigh scattering \citep{goldblatt2009nitrogen}. Second, by increasing the atmospheric mass, the adiabatic lapse rate is increased, that, in turn, warms the surface and lowers the tropopause \citep{goldblatt2009nitrogen}. A third explanation suggested by \cite{chemke2017dynamics} is, that increasing the atmospheric mass, increases the atmospheric heat capacity that in turn weakens the atmospheric radiative cooling. The decrease in radiative cooling, which is more pronounced in colder latitudes, results in warming of the surface and flattening of the meridional temperature gradient. \cite{chemke2017dynamics} also studied the Hadley circulation response to increasing the atmospheric mass, and found that the Hadley circulation narrows and weakens as the atmospheric mass is increased; this is attributed mainly to the lowering of the tropopause and the flattening of the temperature gradient with the increase in atmospheric mass.  

The solar system terrestrial atmospheres of Venus, Earth, Mars, and Titan exhibit significant variability in their different planetary parameters (Table \ref{tab:tab1}), that results in a variety of circulations. For example, Venus has a massive atmosphere, no significant obliquity, and a very slow rotation rate, has a circulation in its lower atmosphere that is composed of two hemispherically symmetric equator to pole Hadley cells, with no seasonality, while its upper atmosphere is dominated by a day to night side circulation \citep[e.g.,][]{read2013dynamics, sanchez2017atmospheric}. On Earth, there is a seasonal cycle, in which the Hadley cell transits from a hemispherically symmetric circulation to a strong, wide, winter cross equatorial cell and a weak and narrow summer cell \citep[e.g.,][]{dima2003seasonality}. 

\begin{table*}[htb!]
\caption{The selected parameters approximated values for the solar system terrestrial atmospheres}
\label{tab:tab1}
\centering
\small
\begin{tabular}{c|cccc}
\hline
\hline
 & Orbital Period & Rotation Period & Obliquity & Atmospheric Mass \\
 \hline
 Venus & $0.6$ years & $243$ days & $177^{\circ}$ & $92$ bar \\
  
 Earth & $1$ years & $1$ day & $23^{\circ}$ & $1$ bar \\

 Mars & $1.8$ years & $1.026$ days & $25^{\circ}$ & $0.006$ bar \\
 
 Titan & $29$ years & $16$ days & $27^{\circ}$ & $1.5$ bar \\
\hline
\end{tabular}
\end{table*}

During Mars' seasonal cycle, its maximum surface temperature shifts from one pole to the other. This strong seasonality is attributed to Mars' thin atmosphere and rocky surface, that also explains Mars' large meridional temperature gradient \citep{mccleese2010structure, read2015physics}. Titan, on the other hand, although having a long orbital period, has a high atmospheric mass and low surface temperature resulting in a long radiative timescale \citep{mitchell2016climate}. As a result, the surface temperature does not shift significantly off the equator during the seasonal cycle \citep{lora2015titan, jennings2016titan}. Also, in contrast to Mars, Titan has an all tropics climate \citep{mitchell2006dynamics}, meaning that its meridional temperature gradient is weak \citep{jennings2009titan}. The low meridional temperature gradient on Titan is attributed mainly to its slow rotation rate \citep{mitchell2016climate, horst2017titan}. The solstice Hadley circulation on both planets is composed of a cross-equatorial cell, with air rising at midlatitudes of the summer hemisphere and descending at midlatitudes of the winter hemisphere \citep[e.g.,][]{lora2015titan, read2015physics}. Although the Hadley cell extent on both planets is similar, on Mars the extent of the circulation is mainly due to the poleward shift of the maximum surface temperature at solstice, and on Titan, it is mainly due to its slow rotation rate \citep{guendelman2018axis}. 

Section \ref{sec:mod} introduces the model used in this study, focusing on the temperature equation. Section \ref{sec:temp} describes how the surface temperature depends on the different parameters. In order to illustrate the effects of the orbital period and atmospheric mass, the case of $\gamma=54^{\circ}$ is studied in detail. Section \ref{sec:stream} describes the zonal mean meridional circulation dependence on the study parameters and the dynamical interpretation of this dependence. We then suggest an empirical power law fit to the model results for the latitudes of the Hadley circulation ascending and descending branches and its strength. Section \ref{sec:conc} summarizes and concludes the results with a discussion on the implications of this study results on possible future observables and planetary habitability.

\section{Model} \label{sec:mod}

In order to study the climate sensitivity to the parameters mentioned above, we use an idealized, aquaplanet, GCM \citep{frierson2007gray}, based on the GFDL dynamical core \citep{gfdl2004new}, similar to the model described in detail in \cite{kaspi2015atmospheric}. The model uses a two-stream gray radiation scheme, where the temperature field is determined by
\begin{equation}
\label{eq:tempeq}
\frac{DT}{Dt}-\frac{R_dT_vw}{c_pp}=Q_r+Q_c+Q_b,
\end{equation}
with the material derivative given by $\frac{D}{Dt}=\frac{\partial}{\partial t}+\mathbf{u\cdot\nabla}$, where $\mathbf{u}=(u,v,w)$ are the velocities in the longitudinal ($\lambda$), meridional ($\phi$) and vertical ($\sigma$) directions. $R_d$ is the dry gas constant of air, $T_v$ is the virtual temperature, which is the temperature that an air parcel would have in a water vapor free air at a given pressure and density. $Q_r, Q_c$ and $Q_b$ are the radiative, convective and boundary layer heating per unit of mass, respectively.

The radiative heating, $Q_r$ is calculated using
\begin{equation}
\label{eq:radheat}
Q_r = \frac{g}{c_p}\frac{\partial}{\partial p}\left(U-D-R_s\right),
\end{equation}  
where $U$ and $D$ are the upward and downward longwave radiation, respectively, and $R_s$ is the solar shortwave radiation. In order to include seasonality, we calculate $R_s$ using
\begin{equation}
R_s=S_0\cos\zeta e^{-\tau_s\sigma},
\end{equation}  
where, $S_0=1360$ Wm$^{-2}$ is the solar constant,  $\tau_s$ is the parameter that controls the vertical absorption of the solar radiation, and $\zeta$ is the diurnal mean zenith angle, which is given by \citep{hartmann2015global, pierrehumbert2010principles}
\begin{equation}
\label{eq:zen}
\cos\zeta=\frac{h}{\pi}\left(\sin\phi\sin\delta+\frac{1}{h}\cos\phi\cos\delta\sin h\right),
\end{equation}
where $\cos h=-\tan\phi\tan\delta$ is the daily average hour angle, where for perpetual night $h=0$ and for perpetual day $h=\pi$. The declination angle is calculated using \citep{pierrehumbert2010principles}
\begin{equation}
\label{eq:declination}
\sin\delta=-\sin\kappa\sin\gamma
\end{equation}
where $\kappa$ is the orbital position relative to the northern hemisphere autumnal equinox angle and $\gamma$ is the obliquity.

Increasing the obliquity, shifts the maximum insolation poleward and increases the meridional insolation gradient (Fig. \ref{fig:figtwo}). For short orbital periods, the yearly mean insolation can become important. The latitude of maximum yearly mean insolation flips from the equator to the poles around obliquity $54^{\circ}$ (Fig.~\ref{fig:figthree}a). Also, the yearly mean meridional insolation gradient is lower for higher obliquities, reaching a minimum at around $\gamma=54^{\circ}$,  (Fig.~\ref{fig:figthree}b). This was shown to be the response of the yearly mean surface temperature \citep{nowajewski2018atmos, kang2019high}.

We use a spectral horizontal resolution of T42 ($2.8^{\circ}\times 2.8^{\circ}$) and $25$ uneven vertical levels for all simulations. The parameters and their range are as follows:
\begin{itemize}
\item Rotation rate, $\Omega=\left[\frac{1}{16},\frac{1}{8},\frac{1}{4},\frac{1}{2},1\right]$.
\item Obliquity, $\gamma=\left[10^{\circ},20^{\circ},30^{\circ},40^{\circ},50^{\circ},60^{\circ},70^{\circ},80^{\circ},90^{\circ}\right]$.
\item Atmospheric mass, $p_s=\left[\frac{1}{2},1 ,2,5,10, 30\right]$.
\item Orbital period, $\omega=\left[\frac{1}{8},\frac{1}{4},\frac{1}{2},1,2,4\right]$.
\end{itemize}
Note that in this study, the units for $\Omega, \ p_s$ and $\omega$ are normalized to the corresponding Earth-like values, where for convenience, we take $\omega=1$ to be $360$ days and not $365$. 

Also note, that in this study, given this vast parameter range, we simulate six different subspaces. We select the subspaces by varying two parameters at a time, holding the other two with Earth-like values, except for the obliquity, that is held at $30^{\circ}$. The range chosen for the different parameters is discussed in section \ref{sec:conc}.

\begin{figure}[htb!]
\centerline{\includegraphics[height=8cm]{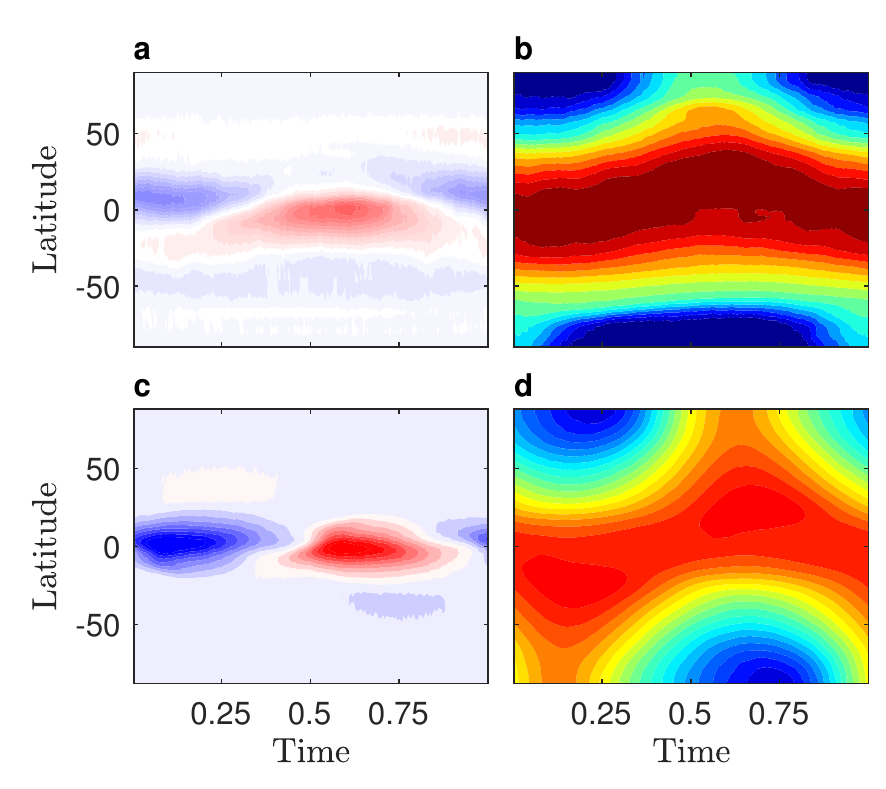}}%{fig_1/pdf_fig/Fig_earth.pdf}}
\caption{A comparison between the Earth (taken from NCEP reanalysis, top row) and the model climate (bottom row). Panels a and c are a Hovmoller diagram of the streamfunction at the height of its time and latitude maximum of the NCEP reanalysis and the model, respectively, with the color scale being $\pm5\times10^7 \rm{m^2 \ s^{-1}}$. Panels b and d are Hovmoller diagrams of the surface temperature of NCEP reanalysis and the model respectively, the color represents $250^{\circ}-300^{\circ}$ K, with red being the warmest. Note, that in all relevant figures in this study, for convenience the time coordinate represents the year fraction.}
\label{fig:fig5}
\end{figure}
 
Comparing the model with an Earth-like configuration and Earth observational reanalysis shows a general similarity between the model results and observations, with some differences. The model exhibits stronger seasonality, a stronger meridional streamfunction, and a colder climate compared with observations (Fig. \ref{fig:fig5}). Taking into account that the model is highly idealized, where for example, clouds, land, or ice effects on the climate are neglected, the resulting climate with Earth-like parameters is satisfying. In addition, as the aim of this study is to examine the climate sensitivity to the planetary parameters to leading order, a perfect reconstruction of Earth's climate is not necessary. Note, that in all relevant figures in this study, for convenience the time coordinate is represented by the year fraction. 

\section{Surface temperature response} \label{sec:temp}
\begin{figure*}[htb!]
\centerline{\includegraphics[height=16cm]{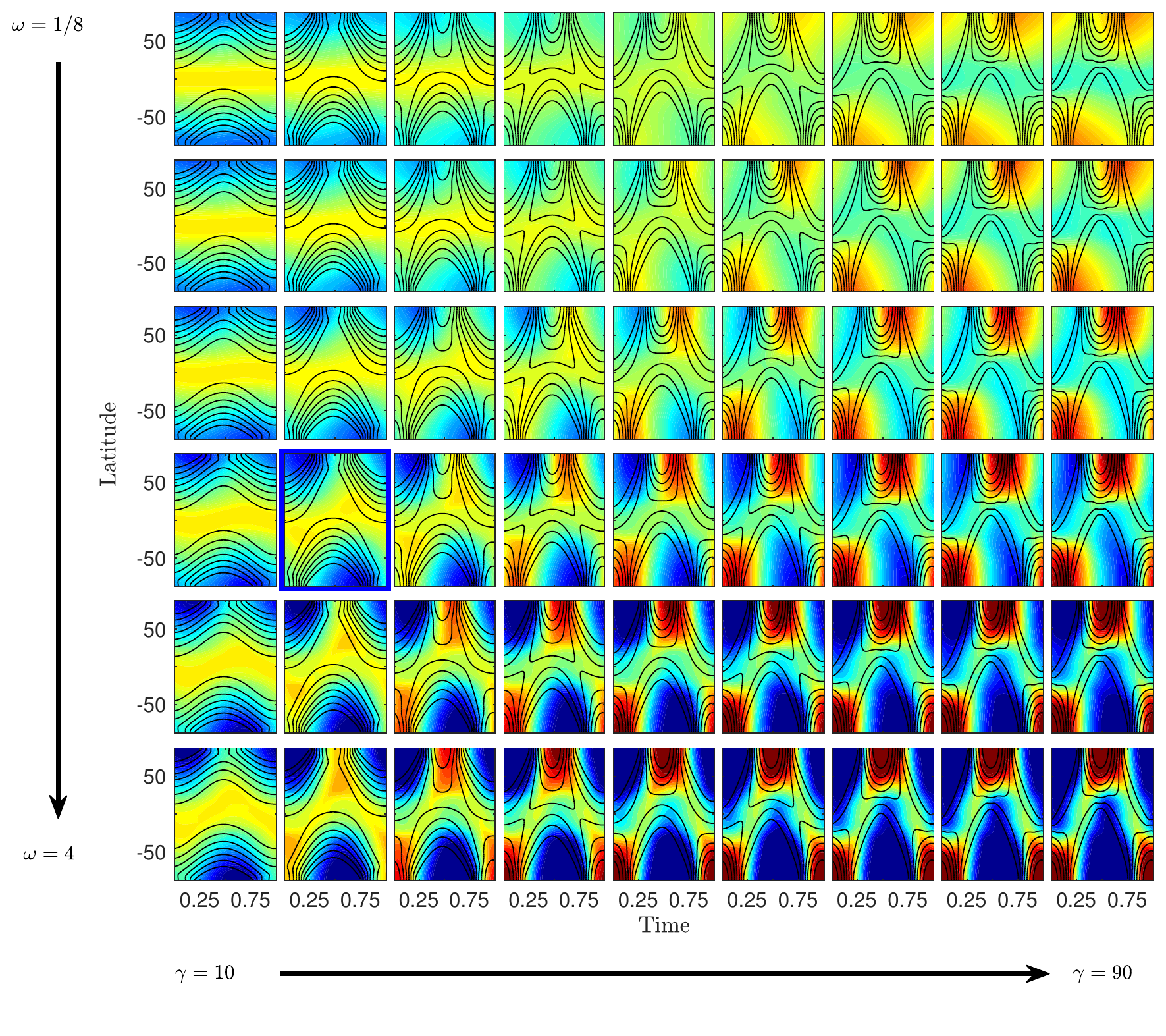}}%{fig_1/pdf_fig/T_t_per_gam.pdf}}
\caption{Hovmoller diagram of the surface temperature (shading, scale is $250-320\rm{K}^{\circ}$, with red being warmest), for different obliquities ($\gamma$) and orbital periods ($\omega$). In all plots rotation rate and atmospheric mass are Earth-like. Black contours represent the insolation. The highlighted panel is for an Earth-like simulation.}
\label{fig:fig6}
\end{figure*}
Visualizing the climate dependence on a broad set of parameters becomes complex when the seasonal cycle is taken into account. A good form to visualize it, are figures the type of Figure \ref{fig:fig6}. This type of figure demonstrates how the climate, and its seasonality, change in response to changes in the parameters in a continuous form. The reader should approach this figure as one plot following the axis that represents the varied parameters. For orientation, we highlight in blue the simulation with Earth-like parameters.

Increasing the obliquity shifts the maximum surface temperature towards the summer hemisphere pole, increases the meridional temperature gradient and results in a sharper transition between summer and winter, where at high obliquities the transition seasons are less pronounced (Fig.~\ref{fig:fig6}). This temperature response to changes in the obliquity is a result of the insolation dependence on the obliquity (Fig. \ref{fig:figtwo}). 

In long orbital periods, the surface temperature follows more closely the solar insolation pattern (black contours in Figure \ref{fig:fig6}). As a result, the climate in long orbital periods exhibits strong seasonality, with maximum surface temperatures reaching to the summer pole (bottom row in Fig. \ref{fig:fig6}). Short orbital periods (top row in Fig. \ref{fig:fig6}), exhibit weak to no seasonality, and the surface temperature takes a similar shape to the yearly mean insolation. The surface temperature dependence on the orbital period is a result of the orbital period being the timescale in which radiative changes occur.This means that long orbital periods allow radiative changes to take place over longer periods, giving the atmosphere more time to adjust to these changes, resulting in a pronounced seasonality. Alternatively, when the orbital period is short, the atmosphere does not have enough time to adjust to the seasonal radiative changes, resulting in the solar forcing being effectively the yearly mean forcing. 
\begin{figure*}[htb!]
\centerline{\includegraphics[height=13cm]{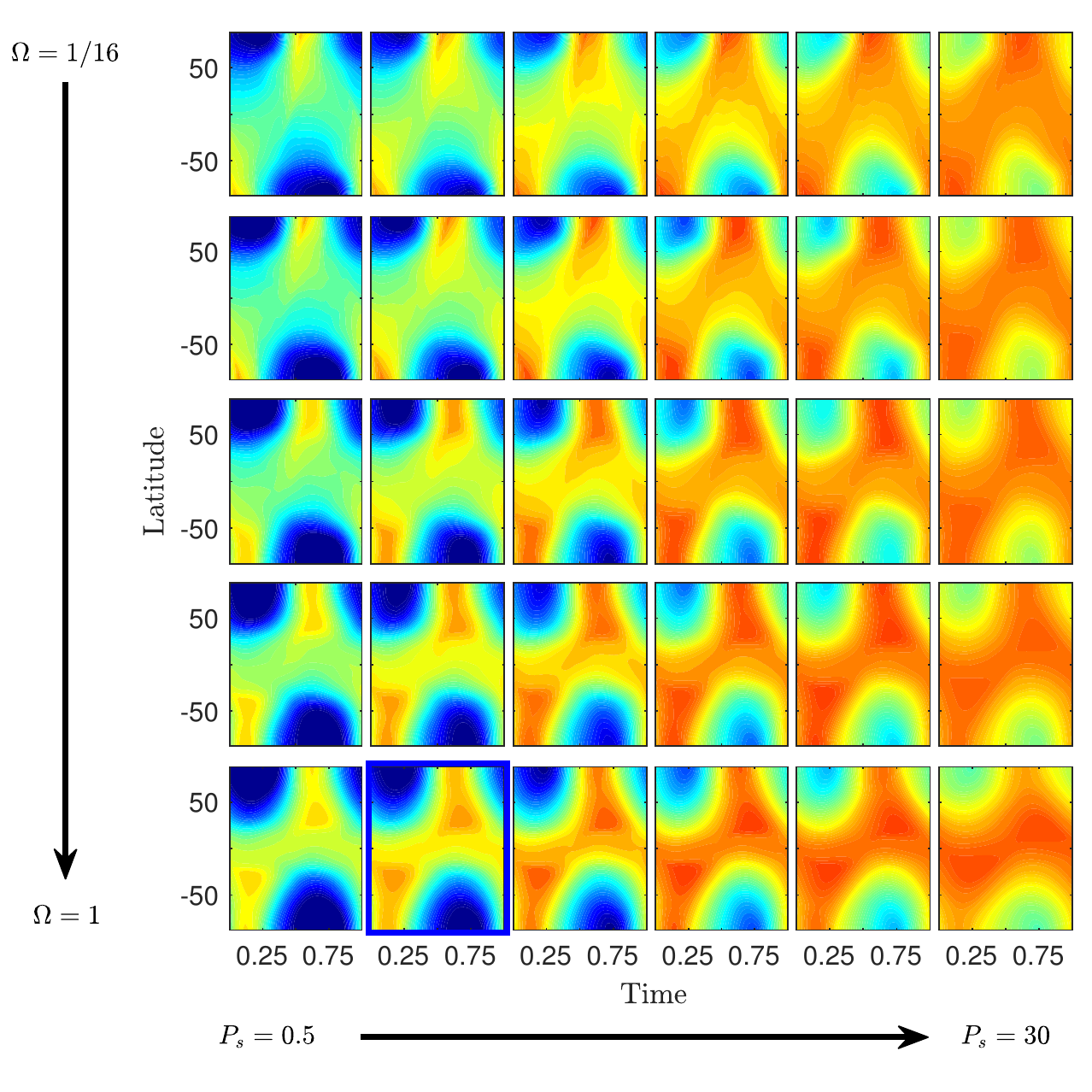}}%{fig_1/pdf_fig/T_t_p_om.pdf}}
\caption{Hovmoller diagram of surface temperature (shading, scale is $260-310^{\circ}\rm{ \ K}$, with red being warmest) for different rotation rates ($\Omega$) and atmospheric masses ($p_s$). Orbital period is Earth-like and $\gamma=30^{\circ}$. Highlighted panel is similar to Figure \ref{fig:fig6}.}
\label{fig:fig7}
\end{figure*}

Faster rotation rates result in a stronger meridional temperature gradient (Fig. \ref{fig:fig7}). This is explained by the eddy scale decreasing with faster rotation rate, which results in a less efficient meridional heat transport, that in turn increases the meridional temperature gradient \citep{walker2006eddy, kaspi2015atmospheric}.

By increasing the atmospheric mass the surface temperature increases and the meridional temperature gradient decreases (Fig. ~\ref{fig:fig7}), in agreement with previous studies \citep[e.g,][]{goldblatt2009nitrogen, chemke2016thermodynamic, chemke2017dynamics}. Using the temperature equation (Equation \ref{eq:tempeq}), \cite{chemke2017dynamics} showed that the main component in the temperature equation that is strongly affected by the atmospheric mass is the radiative term (Equation \ref{eq:radheat}), which can be rewritten as
\begin{equation}
\label{eq:radheatalter}
Q_r = \frac{1}{c_p}\frac{\partial}{\partial m}\left(U-D-R_s\right).
\end{equation}  
Writing $Q_r$ in this form, highlights its dependence on the atmospheric mass ($m=p/g$), or the atmospheric heat capacity, $c_pdm$. Equation \ref{eq:radheatalter} shows that an increase in the atmospheric heat capacity lowers the radiative cooling effect which in turn results in a warmer surface climate and a flatter meridional temperature gradient \citep{chemke2017dynamics}.

Focusing on the atmospheric mass dependence (Fig. \ref{fig:fig7}) shows that the atmospheric mass has only a small effect on the surface temperature seasonality. More specifically, nor the latitude of maximum surface temperature shifts strongly with atmospheric mass neither the temperature temporal variation patterns change significantly, and the main effect of the atmospheric mass, is, as mentioned, a warmer surface climate and flatter meridional temperature gradient. This is puzzling, as the radiative timescale of the atmosphere strongly depends on the atmospheric mass (Equation \ref{eq:radts}).

In order to understand why the atmospheric mass has little effect on seasonality, it is interesting to take a closer look at the $\gamma \approx 54^{\circ}$ case. This is a special case, as it is a turning point in the yearly mean forcing from maximum radiation at the equator to maximum radiation at the poles. Also, it is a minimum point in the yearly mean insolation gradient (Fig. \ref{fig:figthree}). As a result of this unique obliquity value, in a short orbital period, the expected surface temperature will have a relatively small meridional gradient as the effective forcing in a short orbital period resembles the yearly mean forcing (Fig. \ref{fig:fig6}).

\begin{figure}[htb!]
\centerline{\includegraphics[height=8cm]{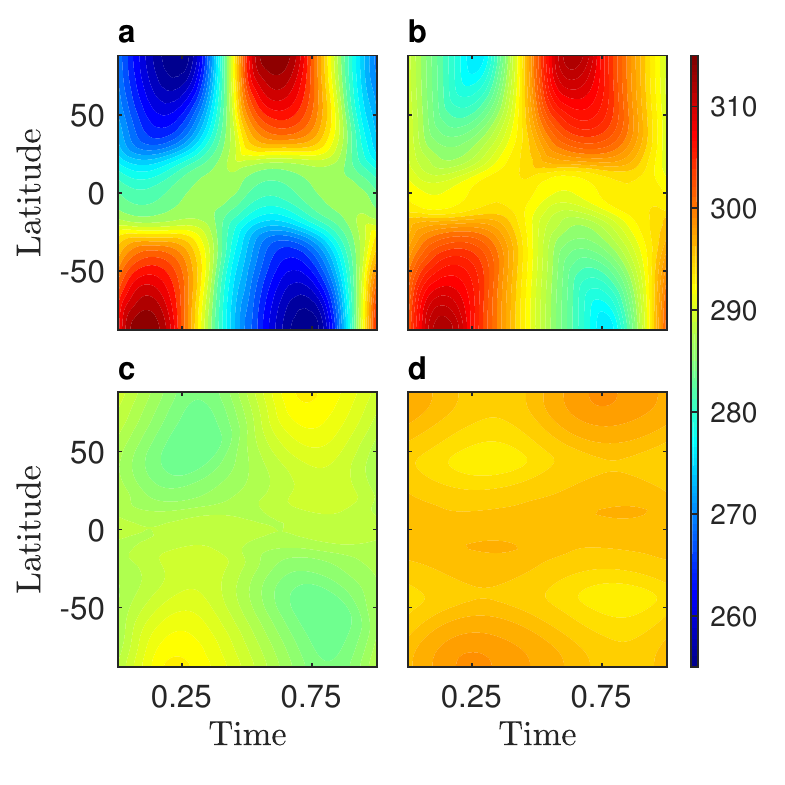}}%{fig_1/pdf_fig/Fig_ob_54.pdf}}
\caption{Hovmoller diagram of surface temperature for different cases of orbital period and atmospheric mass with $\gamma=54^{\circ}$. (a) $\omega$ and $p_s$ are Earth-like, (b) $\omega$ is Earth-like and $p_s=10$, (c) $p_s$ is Earth-like and the $\omega=1/8$, and (d) $p_s=10$ and $\omega=1/8$.}
\label{fig:fig8}
\end{figure} 

\begin{figure}[htb!]
\centerline{\includegraphics[height=4cm]{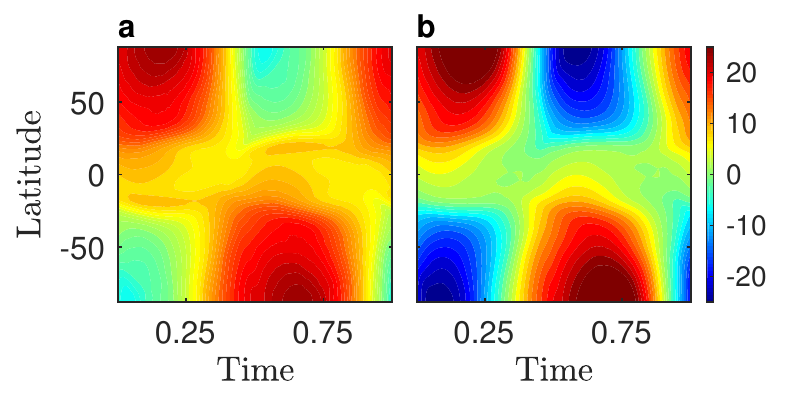}}%{fig_1/pdf_fig/54_diff.pdf}}
\caption{Hovmoller diagram of the surface temperature difference between different cases shown in Figure \ref{fig:fig8}. (a) is the surface temperature difference between $p_s=1$ bar and $p_s=10$ bar both with an Earth-like orbital period ((a) and (b) in Figure \ref{fig:fig8}). (b) is the surface temperature difference between $\omega=1$ and $\omega=1/8$ both with an Earth-like atmospheric mass ((a) and (c) in Figure \ref{fig:fig8}).}
\label{fig:fig9}
\end{figure}

A priori assuming all other parameters stay constant, decreasing the orbital period or increasing the atmospheric mass should have a similar effect on the seasonality as both relate to the adjustment timescale of the atmosphere. For this reason, comparing the effects of increasing the atmospheric mass to decreasing the orbital period, can help to explain why increasing the atmospheric mass has a small effect on seasonality. A quantitative comparison shows that increasing the atmospheric mass mainly warms the surface, with significant warming of the cold latitudes that flatten the meridional temperature gradient (Fig. \ref{fig:fig8}b, Fig. \ref{fig:fig9}a). In contrast, a shorter orbital period tones down the seasonality, by symmetrically decreasing the temperature in warm latitudes and increasing them in cold latitudes (Fig. \ref{fig:fig8}c, Fig. \ref{fig:fig9}b). This asymmetric effect of atmospheric mass on the surface temperature can be explained by the radiative cooling dependence on the atmospheric mass (Equation \ref{eq:radheatalter}). As mentioned, increasing the atmospheric mass decreases the radiative cooling which is more effective at cold latitudes. As a result, there is more warming of cold latitudes than cooling at warm latitudes as the atmospheric mass is increased. This decrease in radiative cooling with the atmospheric mass has a more pronounced effect which diminishes the seasonal effect of the atmospheric mass.

The coupled effect of a short orbital period together with high atmospheric mass results in a climate with close to a latitudinally uniform surface temperature throughout the seasonal cycle for a planet with $\gamma=54^{\circ}$. More specifically, the difference between the yearly maximum and minimum surface temperature is $\sim 6^{\circ}$ K (Fig. \ref{fig:fig8}d). This suggests that not all high obliquity planets will experience a strong seasonal cycle, with fast and strong transitions between warm and cold temperatures, and the possibility that a high obliquity planet will have an equable climate that is habitable exists. It is important to note that diurnal cycle effects are neglected, and although they might become important for short orbital periods, they are out of the scope of this study.  \cite{salameh2017role} showed that diurnal effects on the surface temperature become important when the ratio between the orbital period ($\omega$) and the rotation period ($P_{\Omega}$) is $\omega/P_{\Omega} \approx 10$. In this case, $\omega/P_{\Omega} = 45$, so neglecting diurnal effects is justified.

\begin{figure*}[htb!]
\centerline{\includegraphics[height=8cm]{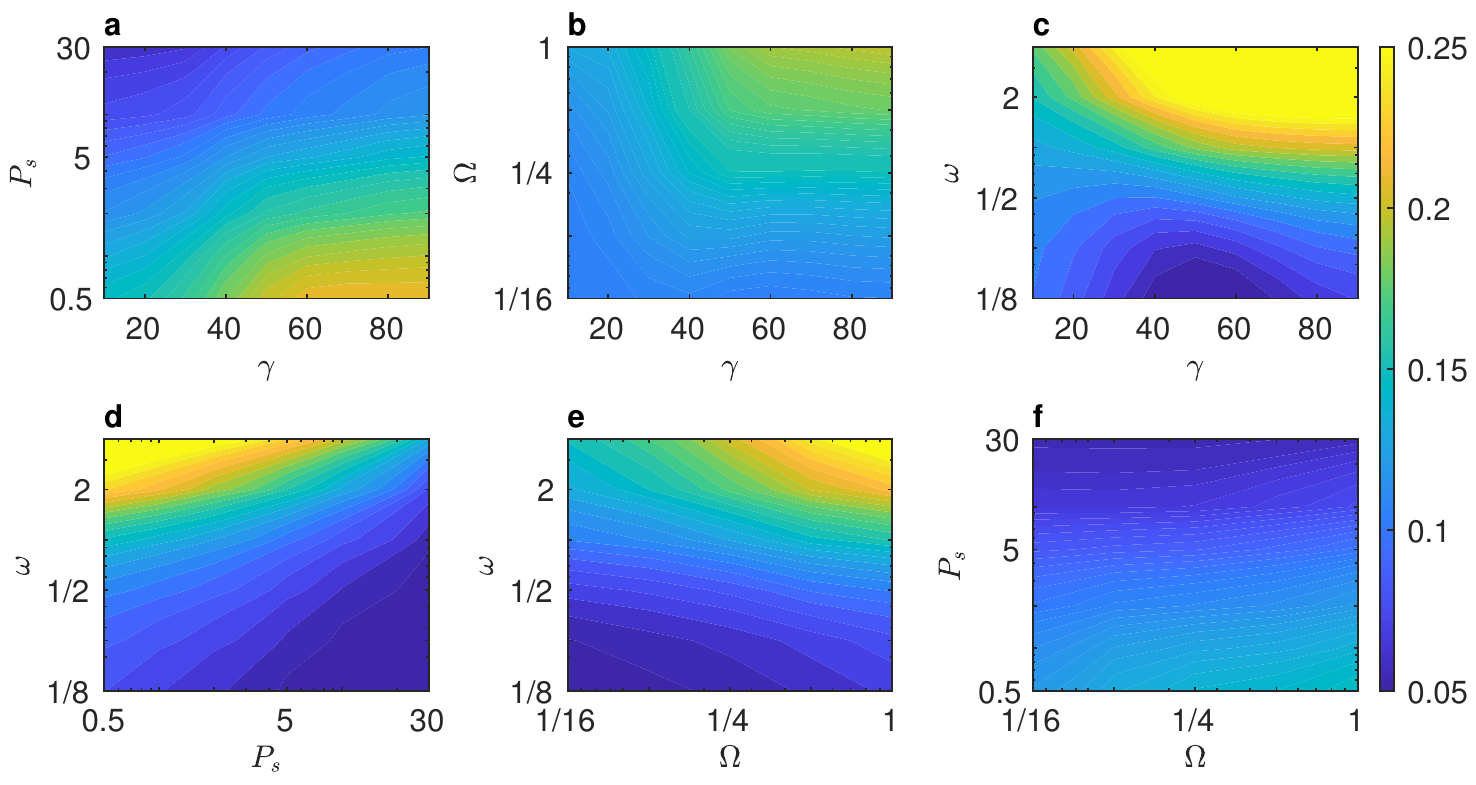}}%{fig_1/pdf_fig/new/del_H.pdf}}
\caption{Normalized meridional surface temperature difference, $\Delta_H=\frac{\rm{max}(\mathit{T_s})-\rm{min}(\mathit{T_s})}{\rm{mean}(\mathit{T_s})}$ for the southern hemisphere summer. Panel a is for atmospheric mass and the obliquity, panel b is for rotation rate and obliquity, panel c is for the orbital period and obliquity, panel d is for orbital period and atmospheric mass, panel e is for orbital period and rotation rate and panel f is for atmospheric mass and rotation rate. The values of the parameters when kept constant are $\gamma=30^{\circ}$, $p_s=1$, $\Omega=1$ and $\omega=1$.}
\label{fig:fig10}
\end{figure*} 

We parametrize the meridional surface temperature difference by 
\begin{equation}
\label{eq:del_H}
\Delta_H=\frac{\rm{max}(\mathit{T_s})-\rm{min}(\mathit{T_s})}{\rm{mean}(\mathit{T_s})},
\end{equation}
where $T_s$ is the surface temperature in the southern hemisphere summer. $\Delta_H$ is a measure for the meridional surface temperature gradient and represents the value of the temperature difference between the warm and cold latitudes in the extreme season. Generally, $\Delta_H$ increases as the rotation rate, obliquity, and orbital period are increased, or when the atmospheric mass is decreased (Fig. \ref{fig:fig10}). However, a clear exception is found for short orbital periods ($\omega<1/4$), were a minimum in $\Delta_H$ is found around obliquity $\sim 50^{\circ}$ (Fig. \ref{fig:fig10}c). This minimum corresponds to the minimum in the yearly mean insolation meridional difference around obliquity $\sim54^{\circ}$ (Fig. \ref{fig:figthree}b), indicating that in short orbital periods, the climate becomes similar to the yearly mean climate.

\section{Hadley circulation response} \label{sec:stream}
\subsection{Simulation results}
The Hadley circulation is a thermally driven component of the zonal mean meridional circulation, with air rising at warm latitudes and descending at colder ones. As air rises, it condenses, creating a zone highly populated with clouds and intense precipitation. On Earth, it is called the intertropical convergence zone (ITCZ). The width, namely the latitude of ascending and descending branches and the seasonality of the Hadley circulation varies between the solar system planets. Venus and Titan are extreme examples, where the mean meridional circulation is composed mainly of the Hadley circulation, which reaches to high latitudes \citep[e.g.,][]{roe2012titan, sanchez2017atmospheric}.  Numerous theories were suggested for predicting the positions of the ascending and descending branch, and a good summary of the different theories can be found in \cite{faulk2017effects}.

The Hadley circulation is closely related to the water cycle on Earth, and the methane cycle on Titan \citep{mitchell2006dynamics, schneider2012titan}. As mentioned, the ascending branch of the circulation correlates to the region of intense precipitation and is a region highly populated with clouds. In contrast, the descending branch region is clear of clouds and generally associated with low precipitation and desert areas. The cloud and desert zones might be good candidates for future observables through their very different albedos. Determining the detection methods for the circulation effects is out of the scope of this study.

\begin{figure*}[htb!]
\centerline{\includegraphics[height=13cm]{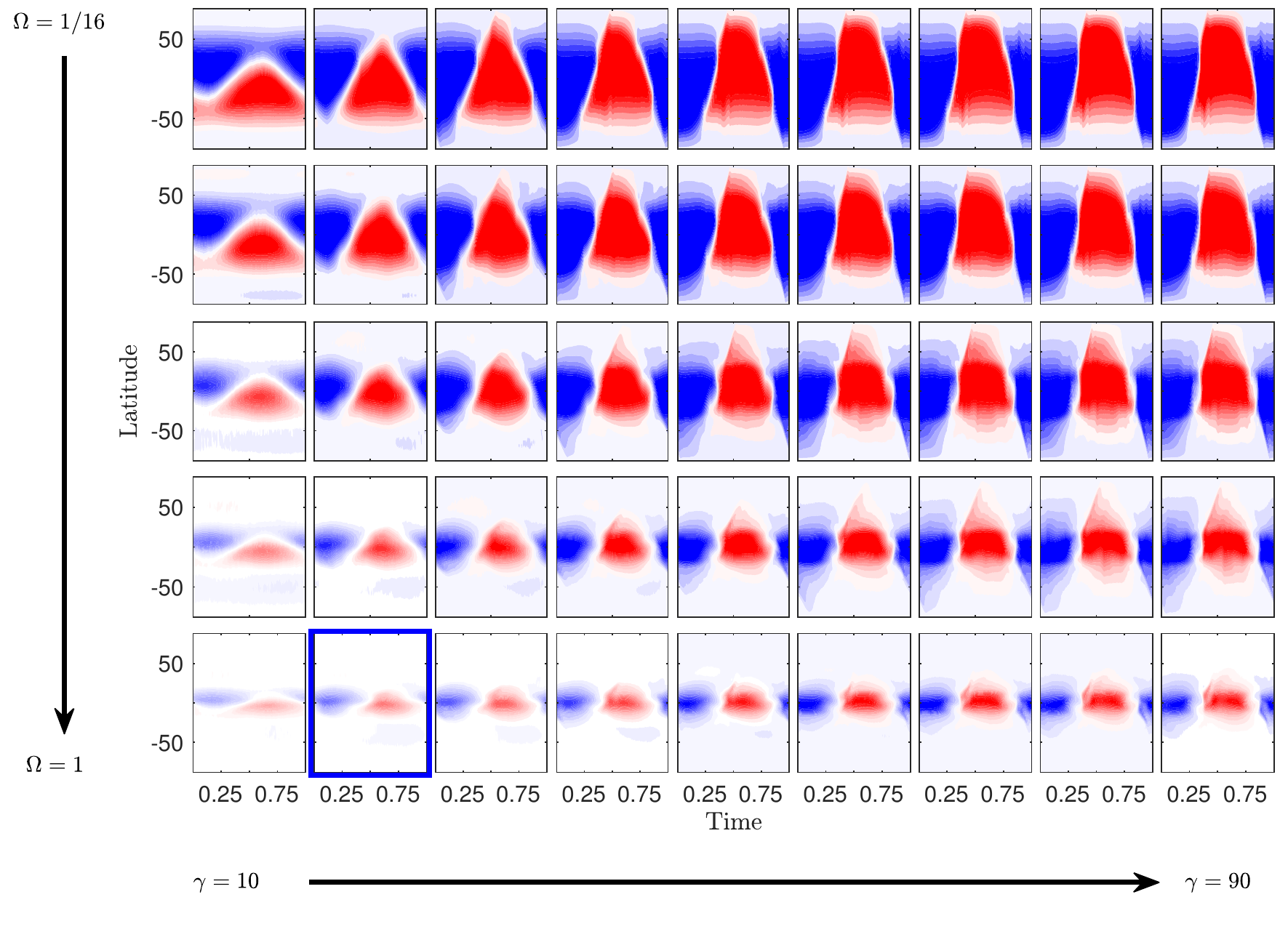}}%{fig_1/pdf_fig/S_t_gam_om.pdf}}
\caption{Hovmoller diagram of the streamfunction (shading, scale is $\pm 1\times10^8 \rm{m^2 \ s^{-1}}$) at its height of maximum (temporal and latitudinal) for different rotation rate and obliquities. Orbital period and atmospheric mass are Earth-like The highlighted panel is the same as in Figure \ref{fig:fig6}.}
\label{fig:fig11}
\end{figure*}

The seasonality of the meridional circulation and its strength increase with increasing obliquity (Fig. \ref{fig:fig11}). During solstice of strong seasonal cases, the zonal mean meridional streamfunction is composed of a strong cross-equatorial winter Hadley cell, with no summer cell and weak transition seasons. This Hadley cell response follows the surface temperature response (Figures \ref{fig:fig6} and \ref{fig:fig11}). However, for Earth-like rotation rate cases, even in cases where the maximum surface temperature is at the pole (fourth row in Figure \ref{fig:fig6}), the ascending branch does not reach the pole (bottom row of Figure \ref{fig:fig11}). This result is in agreement with \cite{faulk2017effects}, who showed that even in an eternal solstice case (for Earth-like forcing), where the maximum surface temperature is at the pole, the ITCZ stays at low latitudes, and does not shift to the position of maximum surface temperature. However, by slowing down the rotation rate, a wider and stronger circulation emerges, extending from the summer hemisphere pole to around latitude $\sim 60^{\circ}$ in the winter hemisphere (Fig. \ref{fig:fig11}). This result indicates that the rotation rate is a limiting factor for the width of the circulation, a result which is consistent with the axisymmetric theory \citep{guendelman2018axis, hill2019axis, singh2019rot}.

\begin{figure*}[htb!]
\centerline{\includegraphics[height=16cm]{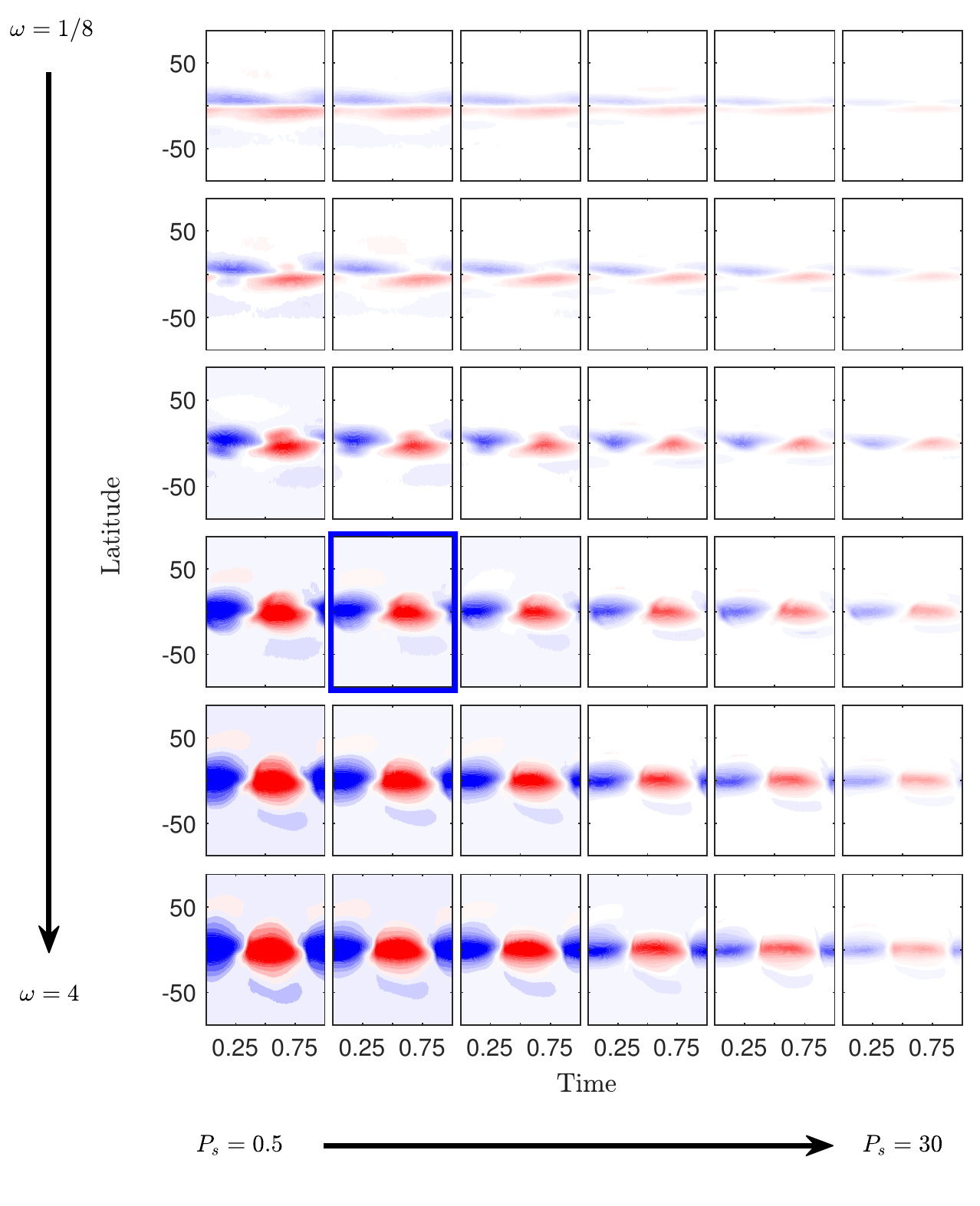}}%{fig_1/pdf_fig/S_t_per_p.pdf}}
\caption{Hovmoller diagram of the streamfunction (shading, scale is $\pm 5\times10^7 \rm{m^2 \ s^{-1}}$) at its height of maximum (temporal and latitudinal) for different orbital periods and atmospheric masses. Rotation is Earth-like and the obliquity is $30^{\circ}$. The highlighted panel is the same as in Figure \ref{fig:fig6}.}
\label{fig:fig12}
\end{figure*}

The circulation narrows and weakens in response to an increase in atmospheric mass, and an opposite response occurs when the orbital period is increased (Fig.~\ref{fig:fig12}). Similar to the response of the circulation to the obliquity changes, the circulation response to the orbital period and atmospheric mass changes also correlates with the surface temperature response (Figures \ref{fig:fig6}, \ref{fig:fig7} and \ref{fig:fig12}). More specifically, increase in seasonality and temperature gradient (where both increases by decreasing the atmospheric mass or increasing the obliquity and orbital period) results in a wider and stronger Hadley circulation (Figures \ref{fig:fig6}, \ref{fig:fig7} and \ref{fig:fig10}--\ref{fig:fig12}). 

\subsection{Theoretical arguments}
Several theories for the ITCZ position and the Hadley cell ascending branch latitude have been suggested. Among them, \cite{neelin1987modeling} theorized that air will ascend close to the latitude of maximum surface temperature or around the maximum of the low level moist static energy (MSE), 
\begin{equation}
m=L_eq+gz+c_pT,
\end{equation}
where $L_e$ is the latent heat of vaporization, $q$ is the specific humidity, and $z$ is the geopotential height. \cite{kang2008response} suggested that the position of the maximum precipitation correlates with the energy flux equator (EFE), which is the latitude where the vertically integrated moist static energy flux ($\overline{mv}$) is zero. 

\begin{figure}[htb!]
\centering
\includegraphics[height=7cm]{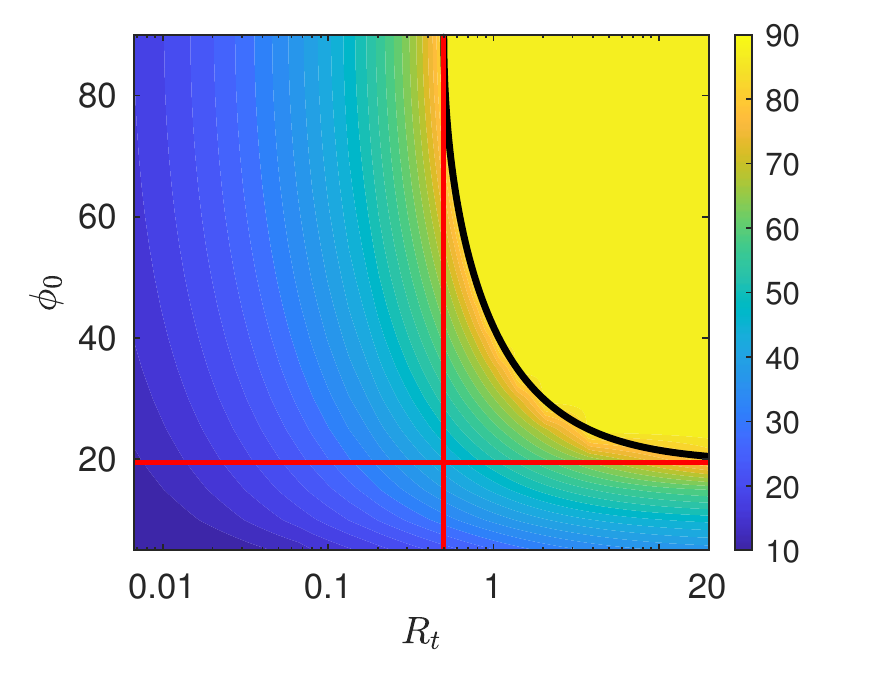}%{fig_1/pdf_fig/axis_Rt}
\caption{Solution of the axisymmetric theory for $\phi_1$, the latitude of the ascending branch (shading, similar to figure 3a in \cite{guendelman2018axis}) for different values of $\phi_0$ (the latitude of maximum radiative equilibrium potential temperature) and $R_t$ (the thermal Rossby number, Equation \ref{thermal_rossby}) . The red lines represent the theoretical limit on a pole-to-pole circulation. The black curve is for Equation \ref{thermal}.}
\label{fig:axis}
\end{figure}

The axisymmetric theory, \citep{held1980nonlinear, lindzen1988hadley, caballero2008axisymmetric, guendelman2018axis}, neglects any eddy contribution to the momentum balance. Assuming angular momentum conservation at the top branch of the circulation, the angular momentum conserving wind is 
\begin{equation}
\label{eq:um}
u_m=\Omega a \frac{\cos^2\phi_1-\cos^2\phi}{\cos\phi},
\end{equation}
where $a$ is the planetary radius and $\phi_1$ is the latitude of the ascending branch (see a more detailed derivation in appendix \ref{append}). Assuming also that gradient-wind balance holds, meaning
\begin{equation}
\label{eq:tmw}
fu+\frac{u^2\tan\phi}{a}=-\frac{1}{a}\frac{\partial \Phi}{\partial\phi},
\end{equation}
with $\Phi=p/\rho_0$, where $p$ is pressure and $\rho_0$ is a reference density used in the Boussinesq approximation \citep{lindzen1988hadley}. Using Equations \ref{eq:um} and \ref{eq:tmw} we can derive the vertical averaged potential temperature field in gradient balance with angular momentum conserving wind \citep{lindzen1988hadley, guendelman2018axis}
\begin{equation}
\label{eq:tempaxis}
\frac{\overline{\theta}(\phi)-\overline{\theta}(\phi_1)}{\theta_0}=-\frac{\delta_H}{R_t}\frac{(\sin^2\phi-\sin^2\phi_1)^2}{\cos^2\phi},
\end{equation}
where $\overline{\theta}$ , the potential temperature, is the temperature that an air parcel would have if it were brought adiabatically to some reference pressure \citep{hartmann2015global}, 
\begin{equation}
\label{thermal_rossby}
R_t=\frac{8\pi\rho GH\delta_H}{3\Omega^2a},
\end{equation}
is the thermal Rossby number \citep{guendelman2018axis}, where $\rho$ is the planet's mean density, $G$ is the universal gravitational constant and $\delta_H$ is the meridional fractional change of the radiative equilibrium temperature that is given by 
\begin{equation}
\frac{\overline{\theta}_e}{\theta_0}=1+\frac{\delta_H}{3}(1-3(\sin\phi-\sin\phi_0)^2),
\end{equation}
where $\phi_0$ is the latitude of maximum $\theta_e$. In order to find the predicted ascending and descending branches of the circulation we need to assume that the circulation is closed and temperature is continuous at the edge of the cells, e.g.,
\begin{eqnarray}
&\int_{\phi_1}^{\phi_j}(\overline{\theta}-\overline{\theta}_e)\cos\phi d\phi &=0, \\
&\overline{\theta}(\phi_j)=\overline{\theta}_e(\phi_j),
\end{eqnarray}
where $j=w,s$ represents the position for the winter and summer descending branch. If we assume that the circulation is composed of only one pole-to-pole cell we can use 
\begin{eqnarray}
&\int_{-\pi/2}^{\pi/2}(\overline{\theta}-\overline{\theta}_e)\cos\phi d\phi =0,\\
&\overline{\theta}(\pi/2) =\overline{\theta}_e(\pi/2),
\end{eqnarray}
to give
\begin{equation}
\label{thermal}
R_t=\frac{1}{3\sin\phi_0 - 1}.
\end{equation}
We can translate Equation \ref{thermal} to conditions on $R_t$ and $\phi_0$ for having a pole-to-pole circulation. First, because $R_t>0$ we need $\phi_0>\arcsin(1/3)$, second, $\phi_0=90^{\circ}$ represents the minimal value of $R_t$ to have a pole-to-pole circulation, meaning that in order to have a pole-to-pole circulation we need $R_t \ge 0.5$. Solving numerically for different values of $R_t$ and $\phi_0$ we see that this condition holds (Fig. \ref{fig:axis}).

This limit on $R_t$ suggests that planets with small $R_t$, (usually fast rotating planets, such as Earth and Mars), will not be able to reach a state with a pole-to-pole Hadley circulation. In contrast, planets with high $R_t$ such as Titan, need a relatively small $\phi_0$ in order to reach a pole-to-pole circulation \citep{guendelman2018axis}.

One advantage of the axisymmetric theory, compared to other theories, is that the axisymmetric theory gives a prediction for both edges of the circulation, ascending and descending, using one set of arguments. However, as mentioned, this theory neglects any eddy contribution, which may play an important role \citep{walker2005response, walker2006eddy}. Other theories that do take into account eddy contribution have been suggested \cite[e.g.,][]{held2000general, korty2008extent, kang2008response, levine2015baroclinic}.

\subsection{Matching theory to simulations} 

\begin{figure*}[htb!]
\centerline{\includegraphics[height=13cm]{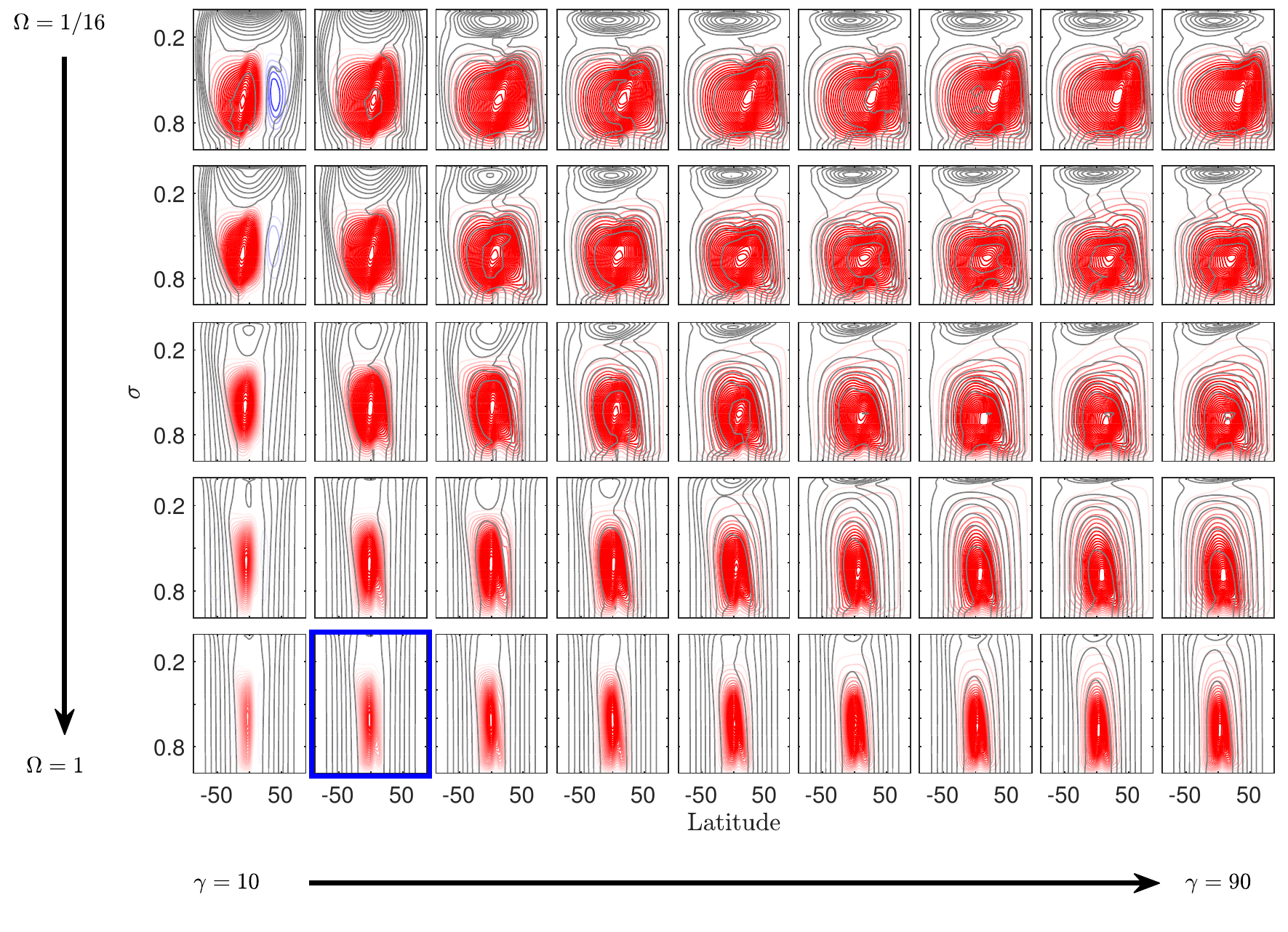}}%{fig_1/pdf_fig/S_z_gam_om_om.pdf}}
\caption{Vertical profile of the streamfunction (colored contours) and zonal mean angular momentum (gray contours) at southern hemisphere summer for different rotation rates and obliquities. Orbital period and atmospheric mass are Earth-like. The highlighted panel is the same as in Figure \ref{fig:fig6}.}
\label{fig:fig13}
\end{figure*}

An alignment of the streamfunction and angular momentum, suggests that the streamfunction conserves its zonal mean angular momentum and the circulation is less influenced by eddies and is more axisymmetric. As a result, cases where the meridional streamfunction follows the zonal mean angular momentum contours, the axisymmetric theory becomes more appropriate. Indeed, in some cases, mainly cases with strong seasonality, an alignment between the streamfunction and zonal mean angular momentum is noticed (Fig. \ref{fig:fig13}).

The streamfunction becomes aligned with the angular momentum contours both by slowing down the planetary rotation rate and by increasing seasonality, namely, having a strong cross-equatorial circulation (Fig. \ref{fig:fig13}). The seasonal alignment is related to a previously studied regime transition from an eddy mediated equinox cell to an axisymmetric solstice cell \citep{bordoni2008monsoons, bordoni2010regime, merlis2013hadley, geen2017regime}. Both alignments are a result of the weakening of eddy momentum flux convergence, where the seasonal alignment due to shielding by the upper level easterlies and the rotation rate alignment is due to a global weakening of the eddy momentum flux convergence as the rotation rate is slowed down \citep{faulk2017effects}.

These alignments, indicate that axisymmetric arguments \citep[e.g.,][]{lindzen1988hadley} can be used to explain the Hadley cell dependence on the different parameters, mainly the circulation dependence on the thermal Rossby number. Note that although this theory is very idealized, it was found to give a good approximation for Earth Hadley cell width  \citep{held1980nonlinear}, for the position of the Hadley cell ascending branch on Titan and Mars \citep{guendelman2018axis}, and can give a physical interpretation to changes in the width and strength of the circulation as a response to changes in planetary parameters \citep{kaspi2015atmospheric, chemke2017dynamics}. As mentioned above, according to the axisymmetric theory, the circulation depends on the thermal Rossby number ($R_t$) and the latitude of maximum warming ($\phi_0$). The theory predicts that the circulation will become wider and stronger as the troposphere height ($H$), the meridional radiative equilibrium temperature gradient ($\delta_H$) and $\phi_0$ are increased or by slowing down the rotation rate \citep[e.g.,][]{guendelman2018axis}. 

Different than the axisymmetric theory, where the radiative equilibrium temperature, more specifically, $\delta_H$, $\phi_0$, and the troposphere height $H$ are predetermined \citep{lindzen1988hadley}, in our model these parameters are not predetermined, so a parameterization using the model results for these parameters is needed. The surface meridional temperature difference $\Delta_H$ (Equation \ref{eq:del_H}) is used to parameterize the meridional radiative equilibrium temperature gradient ($\delta_H$). For the latitude of maximum radiative equilibrium temperature, we choose the latitude of maximum surface temperature. For the troposphere height parameterization, the height of the circulation is used \citep{walker2006eddy}, being the height where the circulation is $10\%$ of its maximum. This parameterization for the tropopause height seems to be a more appropriate one than the WMO lapse rate definition \citep{reichler2003determining}, as the focus of this study is on the circulation itself. Note, that the choice of parameterizations, although with reason, is not unique, and other physically consistent parameterizations can be used.

\begin{figure*}[htb!]
\centerline{\includegraphics[height=8cm]{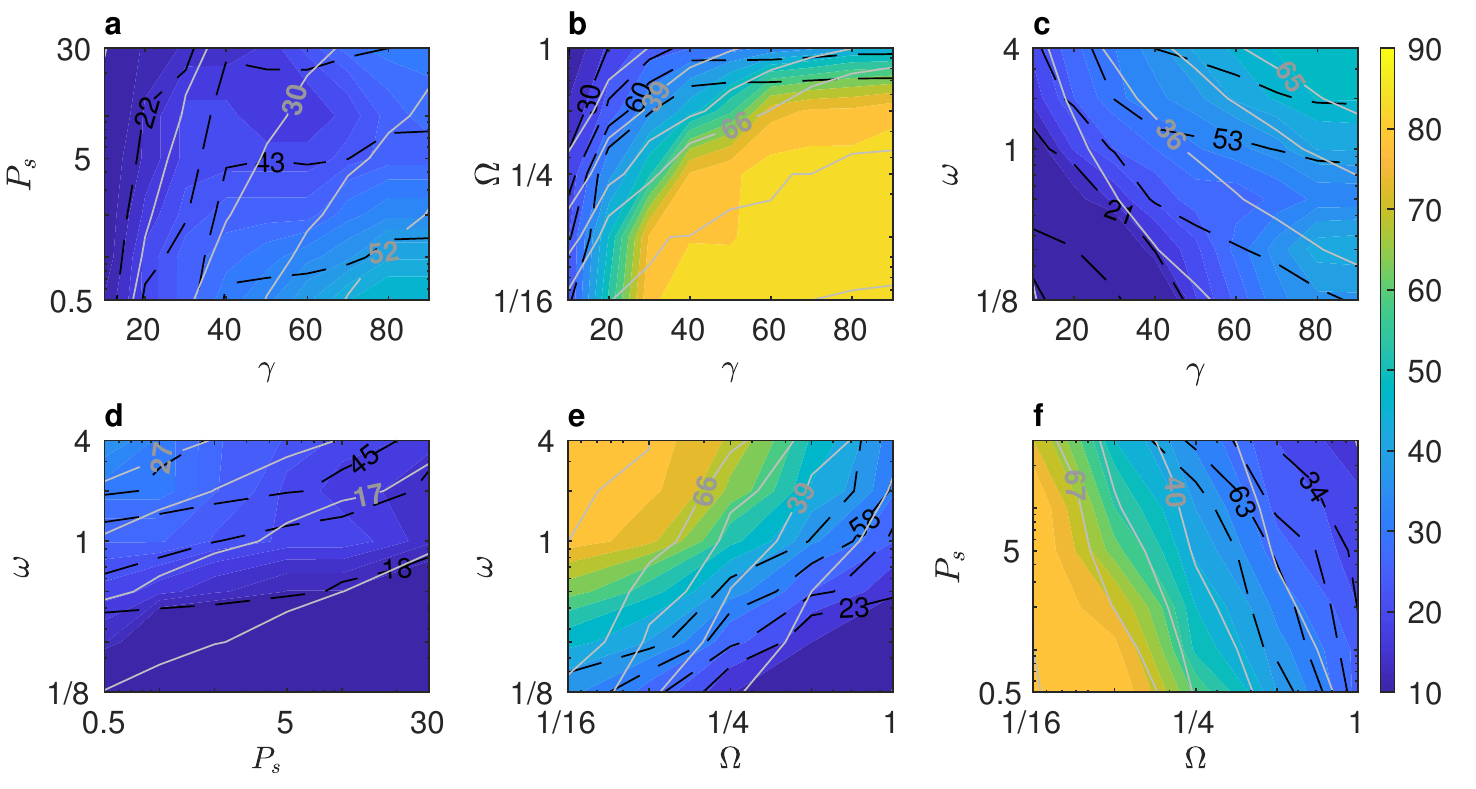}}%{fig_1/pdf_fig/new/phi_1.pdf}}
\caption{The position of the ascending branch $\phi_1$. Shading is for the model results, black dashed contours are for the calculated prediction from the axisymmetric theory and gray contours are for the power law fit. The panel coordinates are the same as in Figure \ref{fig:fig10}.}
\label{fig:figphi1}
\end{figure*} 

Comparing between the model results for the ascending branch (Fig. \ref{fig:figphi1}, shading), taken to be the latitude where the streamfunction gets to $5\%$ of its maximum value \citep{walker2006eddy, faulk2017effects}, with the calculated prediction from the axisymmetric theory (dashed contours in Fig. \ref{fig:figphi1} and Fig.~\ref{fig:figpred}b), shows a general agreement with some numerical misfit. Although the correlation between the axisymmetric prediction and the model results is not perfect, its existence, together with the general agreement between the contour shapes (Fig. \ref{fig:figphi1}) suggests that the use of axisymmetric arguments are appropriate to leading order. Yet, there is a definite numerical misfit, that can be a result of neglecting the eddy contribution. Another possible explanation for this misfit is a problem in the parameterization used to calculate the theoretical prediction, such as the use of the surface temperature to parametrize $\delta_H$. The surface temperature is the final result of the dynamical and radiative processes, so using it to parameterize $\delta_H$ that in the theory context represents only the radiative forcing is problematic, so a separation between the two contributions might be in need for a more appropriate parameterization. 

 \begin{figure*}[htb!]
\centerline{\includegraphics[height=8cm]{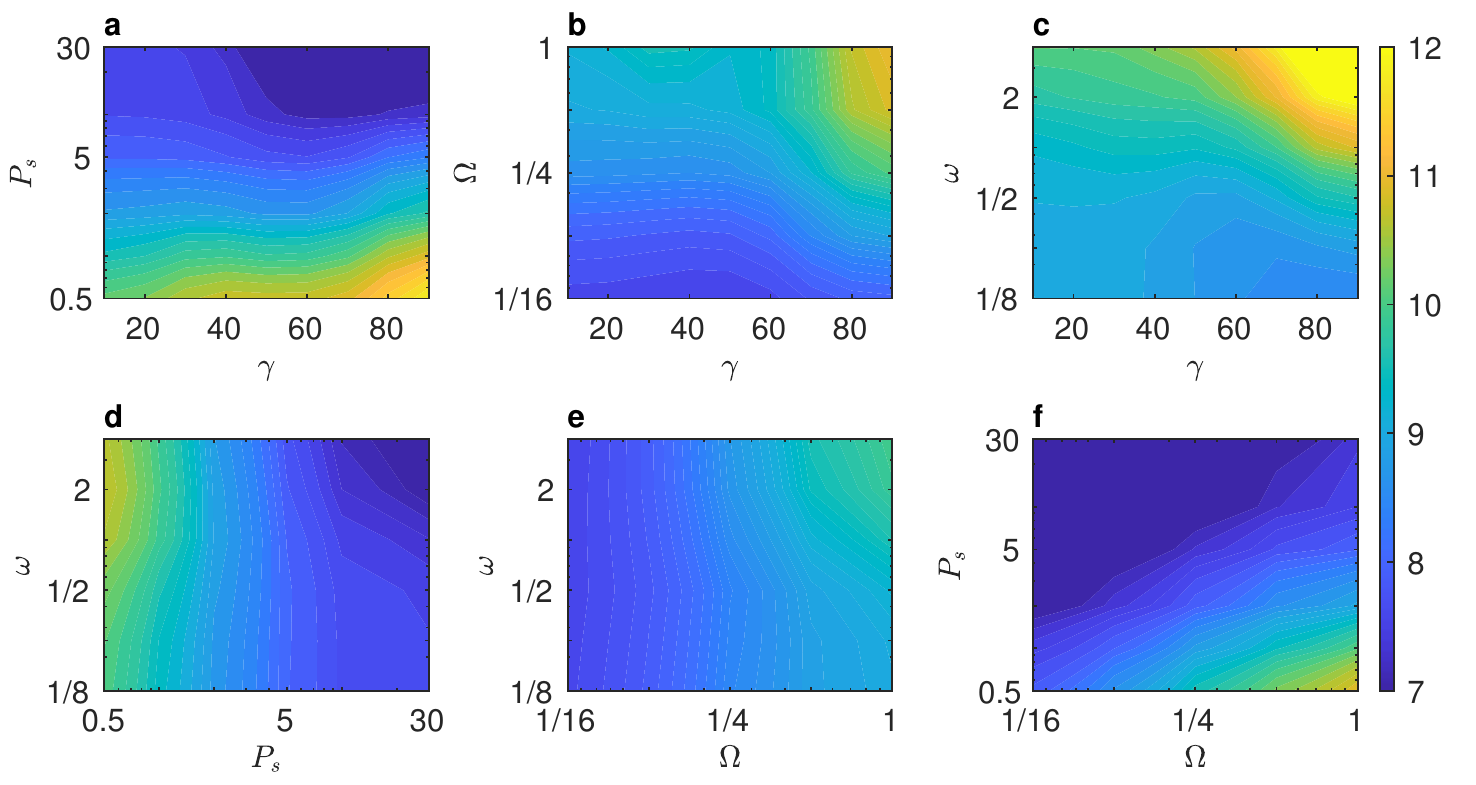}}%{fig_1/pdf_fig/new/H.pdf}}
\caption{The tropopause height as a function of the different parameters in southern hemisphere summer. The panels coordinates are the same as in Figure \ref{fig:fig10}}
\label{fig:figtropo}
\end{figure*}      

Figure \ref{fig:figphi1} summarizes the effect of the different study parameters on the Hadley cell ascending branch latitude. Increasing the orbital period and the obliquity or decreasing the rotation rate and atmospheric mass, results in the poleward shift of the ascending branch. These trends of the ascending branch position with the atmospheric mass, orbital period and obliquity, correlate well with the trends of $H$ and $\Delta_H$. More specifically, increasing the obliquity and orbital period or decreasing the atmospheric mass results in an increase in $H$ and $\Delta_H$ (Figures \ref{fig:fig10} and \ref{fig:figtropo}). The increase in $H$ and $\Delta_H$ correlates with the poleward shift of the ascending branch, in agreement with axisymmetric arguments \citep{held1980nonlinear}. However, increasing the rotation rate increases both the temperature gradient and tropopause height (Figures \ref{fig:fig10} and \ref{fig:figtropo}), and in contrast to the axisymmetric arguments, the circulation narrows. Nonetheless, this is not a violation of the axisymmetric theory; first, because the rotation rate is itself an important parameter in this theory, where by increasing the rotation rate the circulation becomes narrower and this effect dominants over the effect of the other parameters \citep{guendelman2018axis}. Second, the increase in the temperature gradient with faster rotation rate is associated with the decrease of the eddy scale with the rotation rate \citep[e.g.,][]{kaspi2015atmospheric}, which is a dynamical eddy effect; thus it does not contradict axisymmetric arguments.

In agreement with \cite{faulk2017effects} this study suggests that the only form in which the ascending branch can reach the summer pole is by slowing down the rotation rate (Fig. \ref{fig:figphi1}). This strong dependence of the circulation on the rotation rate comes mainly from the dominance of the Coriolis force in the momentum balance. A similar strong dependence on the rotation rate also arises in the axisymmetric theory, where it predicts that the width of the circulation is a function of thermal Rossby number that depends strongly on the rotation rate \citep{guendelman2018axis}.

\begin{figure}[htb!]
\centerline{\includegraphics[height=4cm]{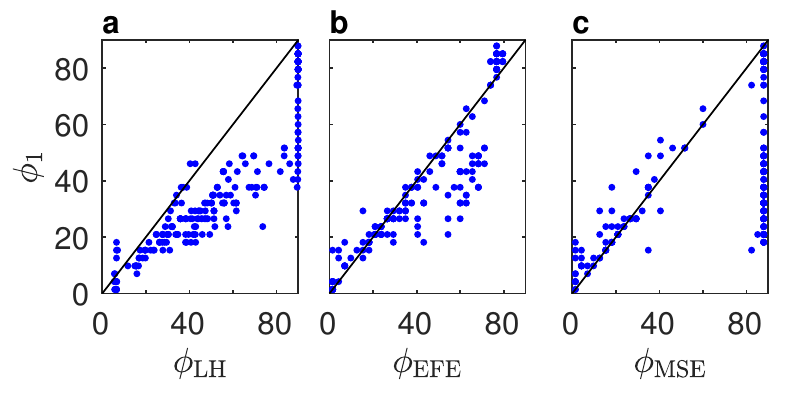}}%{fig_1/pdf_fig/pred.pdf}}
\caption{A comparison between the model results and three chosen theoretical predictors. Panel a is for the axisymmetric predictor \citep{lindzen1988hadley}, panel b is for the energy flux equator predictor \citep{kang2008response} and panel c is for the position of maximum moist static energy predictor \citep{neelin1987modeling}.}
\label{fig:figpred}
\end{figure}  

Comparing the model results with the energy flux equator \citep[e.g.,][]{kang2008response,bischoff2014energetic, schneider2014migrations}, taken here to be the the latitude where the vertically integrated MSE flux ($\overline{mv}$) reaches $5\%$ of its maximum, although somewhat disperse, it shows a good correlation with the position of the ascending branch (Fig.~\ref{fig:figpred}b). These results are in agreement with \cite{wei2018energetic} who showed that there is a lag between the EFE  and the ITCZ seasonal cycle; however, when taking seasonal time average, this lag disappears. 

Another predictor is the latitude of low level maximum MSE \citep[e.g.,][]{neelin1987modeling}, $\phi_{\rm{MSE}}$, where it correlates well with the ascending branch as long as $\phi_{\rm{MSE}}$ stays around low latitudes. However, there are points where $\phi_{\rm{MSE}}$ is at the pole, but the ascending branch stays at low latitudes (Fig.~\ref{fig:figpred}c). These points correspond to cases with relative fast rotation rates (close to Earth-like rotation rate) where the maximum surface temperature is at the pole, but $\phi_1$ does not reach the pole, as the fast rotation rate acts to limit the extension of the circulation \citep{faulk2017effects, guendelman2018axis}.

\subsection{Power law fit for the Hadley circulation response}
In order to get a relation between the circulation response to the studied parameters, we use a power law fit of the form $\propto p_s^a\gamma^b\Omega^c\omega^d$ for the latitude of the ascending branch, descending branch, and the circulation strength. The power law coefficients are determined by minimizing the sum of squared residuals, e.g., the sum of squared differences between the model function and the fitted power law function. For the fit process, all saturated points were left out. 
 
The fit for the ascending branch, $\Phi_1$ is 
\begin{equation}
\label{eq:fitphi1}
\Phi_1\propto p_s^{-0.13}\gamma^{0.71}\Omega^{-0.62}\omega^{0.28},
\end{equation}
with $R^2=0.87$.

\begin{figure}[htb!]
\centerline{\includegraphics[height=6cm]{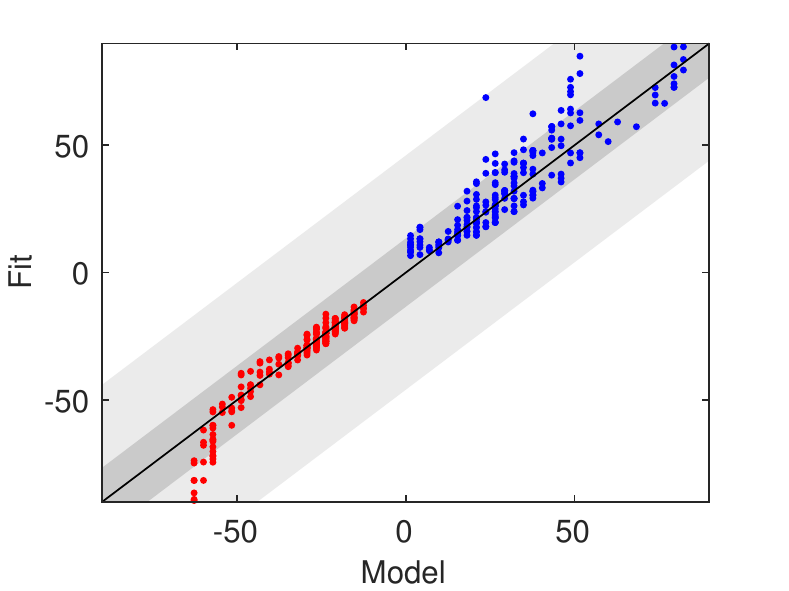}}%{fig_1/pdf_fig/new/fit_phi.pdf}}
\caption{The empirical fit correspondence with the model results, blue dots are for the ascending branch, red dots are for the descending branch, the black line is for a one to one correspondence. Dark gray shading is for $\pm2\sigma$, the light gray shading includes all the points.}
\label{fig:figfitphi}
\end{figure} 

Although possible correlations between the different parameters are ignored, this power law fit does reasonably well as a first order approximation for both the ascending and descending branches (Figures \ref{fig:figphi1} and \ref{fig:figfitphi}). The fit for the descending branch is

\begin{equation}
\label{eq:fitphiH}
\Phi_w\propto p_s^{-0.13}\gamma^{0.14}\Omega^{-0.44}\omega^{0.13},
\end{equation}
with $R^2=0.93$.

\begin{figure}[htb!]
\centerline{\includegraphics[height=6cm]{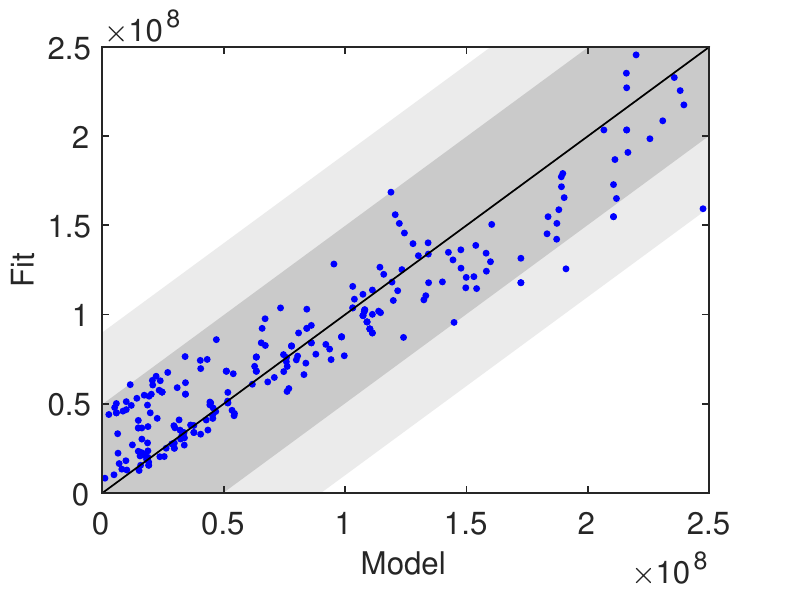}}%{fig_1/pdf_fig/new/fit_m.pdf}}
\caption{The empirical fit correspondence with the model results, for the strength of the circulation, the black line is for a one to one correspondence. Dark gray shading is for $\pm2\sigma$, the light gray shading includes all the points. }
\label{fig:figfitm}
\end{figure}

Another important aspect is the meridional circulation strength (Fig.~\ref{fig:figfitm}), for which the best fit is 
\begin{equation}
\label{eq:fitm}
\Psi_M\propto p_s^{-0.43}\gamma^{0.27}\Omega^{-0.39}\omega^{0.30},
\end{equation}
with $R^2=0.87$.

The first step in studying this empirical fit is to examine the power sign of each parameter. The signs show that the circulation becomes stronger and wider as we increase the obliquity and orbital period (positive sign), and also by decreasing the atmospheric mass and rotation rate (negative sign), in agreement with the results shown in this study. 

The widening and strengthening of the circulation with the obliquity and orbital period correlates with the increase in the seasonality and meridional temperature gradient that results by increasing these parameters (Figures \ref{fig:fig6} and \ref{fig:fig10}). Although there is also some contribution from the increase in the tropopause height with these parameters (Fig. \ref{fig:figtropo}), the tropopause height increase is not as substantial as the increase in the temperature gradient.

Comparing the strength of the circulation and latitude of ascending and descending branch powers of $p_s$  shows a clear difference, where the atmospheric mass strongly affects the circulation strength but has little effect on the width of the circulation. The reduction in the meridional temperature difference (Fig. \ref{fig:fig10}) and of the troposphere height (Fig. \ref{fig:figtropo}) together with the small to no influence on the maximum surface temperature position (Fig.~\ref{fig:fig7}) can explain the difference between the dependence of the circulation strength and width on the atmospheric mass. Our results are consistent with \cite{chemke2017dynamics} that found that in the perpetual equinox case, $\Phi_w\propto p_s^{-0.16}$ and $\Psi_m\propto p_s^{0.41}$.

Previous studies of the Hadley cell dependence on the rotation rate suggested some empirical and theoretical power laws. \cite{caballero2008axisymmetric}, using a small angle approximation showed that the axisymmetric theory predicts that $\Phi_1\propto \Omega^{-2/3}$, which is close to the result presented here for the ascending branch (Equation \ref{eq:fitphi1}). However, \cite{caballero2008axisymmetric} gave the same scaling for $\Phi_w$ which is different from the result here (Equation \ref{eq:fitphiH}), indicating, that the axisymmetric arguments cannot help to fully explain the descending branch position in our simulations and there is a need for other processes. This conclusion is strengthened by the fact that the descending branch in our simulations does not go further than latitude $\sim 60^{\circ}$, which is not a feature of the axisymmetric theory. \cite{faulk2017effects} studied the ITCZ position in the seasonal case, where the insolation was kept Earth-like, and only the rotation rate was changed and found that the position of the ITCZ is proportional to $\Omega^{-0.63}$, in agreement with these results. Other studies gave different theoretical and empirical fits; however, those studies focused mainly on the perpetual equinox case \citep[e.g.,][]{held1980nonlinear, walker2006eddy}.

\section{Conclusion}\label{sec:conc}

The ability to detect and characterize terrestrial exoplanets has improved in the last years, with more planets detected and future missions planned to detect and characterize terrestrial planets and their atmospheres. The planets span over a wide range of orbital configurations, planetary parameters, and atmospheric characteristics. Studying the circulation dependence on different planetary parameters is important in order to understand the climate on these planets and can also give rise to possible future observables.

This study focuses on the climate dependence on four parameters: obliquity, orbital period and atmospheric mass, that strongly relate to the radiative forcing, together with the rotation rate, that has significant dynamical importance. In order to study the climate dependence on these four parameters, we use an idealized GCM, where the simplifications in this model allow to separate the different dynamical effects of each parameter without introducing other complexities such as land, ice or cloud feedbacks. 

The focus is on the zonal mean circulation, specifically the surface temperature and the Hadley circulation. Understanding the surface temperature response to changes in different planetary parameters is important in order to asses the planet's habitability potential. For example, a highly seasonal climate with large and fast transitions between cold to warm temperature reduces the habitability potential of a planet, even if the planet's mean surface temperature is in the habitable range, and surface water can be present. Also, it is essential to understand the water cycle response to changes in the planetary parameters; therefore, we study the Hadley circulation response. The Hadley circulation will also dictate ground features of the planet, for example, on Earth the distinction between the wet tropical regions, located around the ascending branch of the Hadley cell and the deserts located at the descending branch of the Hadley cell. Also on Titan methane lakes around the pole and ice deserts around the equator and midlatitudes can be related to the Hadley circulation \citep[e.g.,][]{Hayes2008, aharonson2009asymmetric}. 

Non-zero obliquity introduces seasonality to the solar insolation, as the obliquity increases towards $90^{\circ}$, the seasonality strengthens and insolation meridional gradient increases (Fig. \ref{fig:figtwo}). When introducing seasonality to the climate system, different timescales that relate to the radiative forcing become relevant. The natural seasonal timescale is the orbital period, which is the time over which radiative changes take place. Short orbital period means that the radiative changes take place over a short amount of time and for a short enough orbital period, the resulting surface temperature is closer to the yearly mean climate. Long orbital periods result in a longer time for radiative changes to take place. As a result, the surface temperature in long orbital periods will have a similar structure to the insolation (Fig. \ref{fig:fig6}). 

The atmospheric radiative adjustment timescale, the time that the atmosphere needs to adjust to changes in the radiative forcing, depends on the atmospheric mass (Equation \ref{eq:radts}). However, increasing the atmospheric mass also decreases the radiative cooling which is a dominant effect on the temperature, and it masks over the seasonal effect of atmospheric mass, and as a result, the seasonality is weakly affected by changes in the atmospheric mass (Figures \ref{fig:fig7} and \ref{fig:fig9}). That being said, we can still see some effect of the atmospheric mass on the seasonality, more specifically, the equatorward shift of the maximum surface temperature for high enough atmospheric mass (Fig. \ref{fig:fig7}). The $\gamma\approx 54^{\circ}$ case is an interesting one, mainly because it exhibits a minimum in the yearly mean insolation (Fig. \ref{fig:figthree}). As a result of this minimum, in a short orbital period and a high atmospheric mass planet, the climate becomes equable, with no abrupt seasonal transitions (Fig. \ref{fig:fig8}d), this can increase the planets' habitability potential.

The response of the circulation to changes in obliquity, orbital period and atmospheric mass follow the response of the temperature. Meaning that by increasing the obliquity and orbital period, the circulation becomes wider and stronger (Figures \ref{fig:fig11} and \ref{fig:fig12}) due to the increase in seasonality and the meridional temperature gradient (Fig. \ref{fig:fig10}). Decreasing the atmospheric mass also results in a wider and stronger circulation (Fig.~\ref{fig:fig12}), attributed mainly to the increase in the meridional temperature gradient and the increase in the troposphere height \citep{chemke2017dynamics}. The effect of the atmospheric mass on the circulation strength is stronger than the effect on its width (Equations \ref{eq:fitphi1}--\ref{eq:fitm}). This also correlates to the temperature response, as mentioned, the atmospheric mass does not contribute significantly to seasonality changes, but does change the meridional temperature gradient (Fig.~\ref{fig:fig10}) and tropopause height (Fig.~\ref{fig:figtropo}).

It is important to note that even in a very seasonal case, where the maximum surface temperature is at the summer pole, the circulation does not extend to the pole if the rotation rate is Earth-like (Fig.~\ref{fig:fig11}), and only happens when the rotation rate is reduced \citep{guendelman2018axis}. Slowing down the rotation rate also flattens the temperature gradient due to the higher heat transport efficiency of larger eddies. In addition, lowering the rotation rate widens the region where the weak temperature gradient limit holds, i.e., where the Coriolis term is less dominant, and gravity waves are acting to smooth the temperature gradient, resulting in a wider tropics \citep{sobel2001wtg, raymond2005wtg}. The widening of the circulation by lowering the rotation rate can be attributed to the decrease in the eddy momentum flux convergence with the rotation rate \citep{faulk2017effects}.

Summarizing the circulation dependence on the different parameters, we suggest an empirical power law fit for the ascending and descending branches and the circulation strength (Equations \ref{eq:fitphi1}, \ref{eq:fitphiH}, \ref{eq:fitm}, respectively). Although this fit is for a specific model in a specific configuration, it gives a first order approximation for the dependence of the circulation on these four parameters. 

Although theoretically the range of the orbital period, rotation rate and atmospheric mass can be even larger, the range that this study covers represents the different seasonal regimes of the diurnal mean climate. The chosen values for the orbital period covers climates raging from close to an annual mean climate for the short orbital periods, to a strong seasonal climate where the surface temperature follows closely the insolation (Fig.~\ref{fig:fig6}). The atmospheric mass range shows both the strong radiative cooling and the relatively weak radiative timescale effects (Fig.~\ref{fig:fig7}). For the rotation rate, we cover mainly the range for Earth-like and slower rotation rate, which covers the circulation response from a narrow, hemispherically symmetric Hadley circulation to a wide cross equatorial circulation (Fig.~\ref{fig:fig11}). Faster rotation rates were discarded as it enters to a different dynamical regime \citep{kaspi2015atmospheric, chemke2015poleward,chemke2015latitudinal}.

In future observations, with some spatial resolution or other ways of inferring cloud-cover, regions highly populated with clouds can possibly be associated with the ascending branch of the Hadley cell, and a sector that is cloud depleted can be related to the descending branch of the Hadley cell. Detecting such temporal changes in these cloud regions latitude can give an indication about the planetary orbital and atmospheric characteristics.

\acknowledgments

We thank Rei Chemke for his help in the model configuration. We also appreciate the helpful comments from the anonymous reviewer. We acknowledge support from the Israeli Science foundation (grant 1819/16) and the Helen Kimmel Center for Planetary Science at the Weizmann Institute of Science.  

\appendix 
%\begin{appendices}
\section{The axisymmetric theory derivation}
\label{append}
Following closely the derivation in \cite{lindzen1988hadley}, we start with the zonal mean angular momentum per unit mass, defined to be
\begin{equation}
M=(\Omega a^2 \cos\phi+ua)\cos\phi,
\end{equation}
where $\Omega$ is the planetary rotation rate, $a$ is the planetary radius $\phi$ is latitude and $u$ is the zonal mean zonal wind. The angular momentum of air that starts at rest from latitude  $\phi_1$, the Hadley circulation ascending branch, is
\begin{equation}
M=\Omega a^2\cos^2\phi_1.
\end{equation}
Assuming that the upper branch of the Hadley circulation conserves its angular momentum, we can write an expression for the angular momentum conserving wind,
\begin{equation}
\label{eq:aum}
u_m=\Omega a\frac{\cos^2\phi_1-\cos^2\phi}{\cos\phi}.
\end{equation}
Assuming also that the zonal flow is in gradient-wind balance
\begin{equation}
\label{eq:athw}
fu+\frac{u^2\tan\phi}{a}=-\frac{1}{a}\frac{\partial \Phi}{\partial\phi},
\end{equation}
where $\Phi=p/\rho_0$, we can evaluate this balance (Equation \ref{eq:athw}) at heights $H$ and $0$ and subtract the two to give
\begin{equation}
\label{eq:athw2}
fu(H)+\frac{u(H)^2\tan\phi}{a}=-\frac{1}{\rho_0 a}\frac{\partial }{\partial\phi}\left[\Phi(H)-\Phi(0)\right],
\end{equation}
where we assume that $|u(0)|\ll|u(H)|$. Assuming hydrostatic balance
\begin{equation}
\label{eq:ahyd}
\frac{\partial\Phi}{\partial z} = \frac{g}{\theta_0}\theta,
\end{equation}
where $\theta$ is the potential temperature and $\theta_0$ is some reference potential temperature, and integrating the hydrostatic balance (Equation \ref{eq:ahyd}) with respect to $z$ gives
\begin{equation}
\label{eq:ahyd2}
\left[\Phi(H)-\Phi(0)\right]=\frac{g}{\theta_0}\overline{\theta},
\end{equation}
where $\overline{\theta}=\int_0^H\theta dz/H$. Substituting $u(H)=u_m$ and Equation \ref{eq:ahyd2} into Equation \ref{eq:athw2} gives
\begin{equation}
\label{eq:atwb}
\frac{1}{\theta_0}\frac{\partial\overline{\theta}}{\partial\phi}=\frac{\Omega^2a^2}{gH}\left(\frac{\tan\phi}{\cos^2\phi}\cos^4\phi_1-\sin\phi\cos\phi\right).
\end{equation}
Integrating Equation \ref{eq:atwb} with respect to $\phi$ yields
\begin{equation}
\label{eq:axisapp}
\frac{\overline{\theta}(\phi)-\overline{\theta}(\phi_1)}{\theta_0}=-\frac{\Omega^2a^2}{2gH}\frac{(\sin^2\phi-\sin^2\phi_1)^2}{\cos^2\phi},
\end{equation}
which is similar to Equation \ref{eq:tempaxis}. In order to get the exact expression as in Equation \ref{eq:tempaxis}, one needs to use $g=4\pi G\rho a/3$ where $\rho$ is the planet's mean density and $G$ is the universal gravitational constant and the definition for the thermal Rossby number 
\begin{equation}
\label{eq:arosb}
R_t=\frac{8\pi\rho GH\delta_H}{3\Omega^2a},
\end{equation}
where $\delta_h$ is meridional fractional change of the radiative equilibrium (Equation \ref{eq:atr}). Substituting $R_t$ in Equation \ref{eq:axisapp} gives Equation \ref{eq:tempaxis} exactly. 

To close the set of equations, we parameterize the thermal forcing using Newtonian relaxation to an equilibrium potential temperature of the form
\begin{equation}
\label{eq:atr}
\frac{\overline{\theta}_e}{\theta_0}=1+\frac{\delta_H}{3}(1-3(\sin\phi-\sin\phi_0)^2),
\end{equation}
with $\phi_0$ being the latitude of maximum $\overline{\theta}_e$. In order to find the predicted latitude of ascending and descending branches of the circulation we need to assume that the circulation is thermally closed and that the temperature is continuous at the edge of the cells, e.g.,
\begin{eqnarray}
\label{A1}
&\int_{\phi_1}^{\phi_j}(\overline{\theta}-\overline{\theta}_e)\cos\phi d\phi =0, \\
\label{A2}
&\overline{\theta}(\phi_j) =\overline{\theta}_e(\phi_j),
\end{eqnarray}
where $j=w,s$ represents the position for the winter and summer descending branch respectively.

Of particular interest is the pole-to-pole circulation case. In this scenario Equations \ref{A1} and \ref{A2} translate to 
\begin{eqnarray}
&\int_{-\pi/2}^{\pi/2}(\overline{\theta}-\overline{\theta}_e)\cos\phi d\phi =0, \\
&\overline{\theta}(\pi/2) =\overline{\theta}_e(\pi/2),
\end{eqnarray}
writing these equations explicitly, with $\phi_1=\pi/2$ gives
\begin{eqnarray}
&\int_{-\pi/2}^{\pi/2}\left[1+\frac{\delta_H}{3}(1-3(\sin\phi-\sin\phi_0)^2)\right]\cos\phi d\phi = \int_{-\pi/2}^{\pi/2}\left[\frac{\theta(\pi/2)}{\theta_0}-\frac{\delta_H}{R_t}\frac{(\sin^2\phi-1)^2}{\cos^2\phi}\right]\cos\phi d\phi,\\
&\frac{\overline{\theta}(\pi/2)}{\theta_0} =1+\frac{\delta_H}{3}(1-3(1-\sin\phi_0)^2).
\end{eqnarray}
After some simple algebraic manipulations of these equations, one can obtain the expression given in Equation \ref{thermal}.

\listofchanges

\end{document}